\documentclass[structabstract]{aa}

\usepackage{txfonts}
\usepackage{natbib}
\bibpunct{(}{)}{;}{a}{}{,} 
\usepackage{graphicx}
\usepackage{deluxetable}

\begin{document}

\title{Dense molecular cocoons in the massive protocluster W3~IRS5: a
  test case for models of massive star formation}

\author{K.-S. Wang\inst{1} \and T. L. Bourke\inst{2} \and
  M. R. Hogerheijde\inst{1} \and F. F. S. van der Tak\inst{3,4} \and
  A. O. Benz\inst{5} \and S. T. Megeath\inst{6} \and
  T. L. Wilson\inst{7}}

\offprints{K.-S. Wang, \email{kswang@strw.leidenuniv.nl}}

\institute{Leiden Observatory, Leiden University, P.O. Box 9513, 2300
  RA Leiden, The Netherlands; email: kswang@strw.leidenuniv.nl \and
  Harvard-Smithsonian Center for Astrophysics, 60 Garden Street,
  Cambridge, MA 02138, USA \and SRON Netherlands Institute for Space
  Research, Landleven 12, 9747 AD, Groningen, The Netherlands \and
  Kapteyn Astronomical Institute, University of Groningen, The
  Netherlands \and Institute of Astronomy, ETH Z\"{u}rich, 8093
  Z\"{u}rich, Switzerland \and Ritter Observatory, MS-113, University
  of Toledo, 2801 W. Bancroft St., Toledo, OH 43606, USA \and Naval
  Research Laboratory, Code 7210, Washington, DC 20375, USA}

\date{Received/Accepted}

\abstract 
{Two competing models describe the formation of massive stars in
  objects like the Orion Trapezium. In the turbulent core accretion
  model, the resulting stellar masses are directly related to the mass
  distribution of the cloud condensations. In the competitive
  accretion model, the gravitational potential of the protocluster
  captures gas from the surrounding cloud for which the individual
  cluster members compete.}
{With high resolution submillimeter observations of the structure, kinematics, and
  chemistry of the proto-Trapezium cluster W3~IRS5, we aim to
  determine which mode of star formation dominates.}
{We present 354~GHz Submillimeter Array observations at resolutions of
  $1\arcsec$--$3\arcsec$ (1800--5400 AU) of W3~IRS5. The dust
  continuum traces the compact source structure and masses of the
  individual cores, while molecular lines of CS, SO, SO$_2$, HCN,
  H$_2$CS, HNCO, and CH$_3$OH (and isotopologues) reveal the gas
  kinematics, density, and temperature.}
{The observations show five emission peaks (SMM1--5). SMM1 and SMM2
  contain massive embedded stars ($\sim$20 $M_{\sun}$); SMM3--5 are
  starless or contain low-mass stars ($<$8 $M_{\sun}$). The inferred
  densities are high, $\ge10^{7}$ cm$^{-3}$, but the core masses are
  small, $0.2-0.6$ $M_{\sun}$. The detected molecular emission reveals
  four different chemical zones. Abundant ($X\sim$ few $10^{-7}$ to
  $10^{-6}$) SO and SO$_2$ are associated with SMM1 and SMM2,
  indicating active sulfur chemistry. A low abundance
  ($5\times10^{-8}$) of CH$_3$OH concentrated on SMM3/4 suggest the
  presence of a hot core that is only just turning on, possibly by
  external feedback from SMM1/2. The gas kinematics are complex with
  contributions from a near pole-on outflow traced by CS, SO, and HCN;
  rotation in SO$_2$, and a jet in vibrationally excited HCN.}
{The proto-Trapezium cluster W3~IRS5 is an ideal test case to
  discriminate between models of massive star formation. Either the massive stars accrete locally from their local cores; in this case the small core masses imply that W3 IRS5 is at the very end stages (1000 yr) of infall and accretion, or the stars are accreting from the global collapse of a massive, cluster forming core.  We find that the observed masses, densities and line widths observed toward W3 IRS 5 and the surrounding cluster forming core are consistent with the competitive accretion of gas at rates of $\dot{M}$ $\sim10^{-4}$ $M_{\sun}$ yr$^{-1}$ by the massive young forming stars. Future mapping of the gas kinematics from large to small
  scales will determine whether large-scale gas inflow occurs and how
  the cluster members compete to accrete this material.}


\titlerunning{SMA observations of W3 IRS5}
\maketitle 

\section{Introduction}
In contrast to low-mass star formation, there is no single agreed upon
scenario for the formation of massive stars
\citep{Zinnecker07}. Current models either predict that a turbulent
core collapses into a cluster of stars of different mass
\citep[e.g.][]{McKee03}, or that multiple low-mass stars compete for
the same mass reservoir, with a few winning and growing to become
massive stars while most others remain low-mass stars
\citep[e.g.][]{Bonnell01,Bonnell06}. Both theories are supported by observations
\citep[e.g.][]{Cesaroni97,Pillai11}, suggesting that there may be two
different modes of massive star formation \citep{Krumholz09}. However,
it is challenging to observe massive star-forming regions since they
are at large distances (a few kpc), requiring high-angular resolution
to resolve the highly embedded complex structures in which
gravitational fragmentation, powerful outflows and stellar winds, and
ionizing radiation fields play influence the star-formation processes
\citep{Beuther07}.

Although observationally challenging, progress in testing the models
of massive star formation has been made recently with at centimeter,
and (sub)millimeter wavelengths. Based on observations of the
protocluster NGC 2264 with the IRAM 30m telescope, \citet{Peretto06}
proposed a mixed type of star formation as proposed by
\citet{Bonnell01} and \citet{Bonnell06}, and \citet{McKee03}, where the turbulent, massive
star-forming core is formed by the gravitational merger of lower mass
cores at the center of a collapsing protocluster. Observations of
G8.68--0.37 with the Submillimeter Array (SMA) and the Australia
Telescope Compact Array (ATCA) imply that to form an O star, the
star-forming core must continuously gain mass by accretion from a
larger mass reservoir since protostellar heating from the low-mass
stars is not sufficient to halt fragmentation
\citep{Longmore11}. Multi-wavelength observations of G29.96--0.02 and
G35.20--1.74 suggest that the mass of massive star-forming cores is
not limited to the natal cores formed by fragmentation and turbulence
is not strong enough to prevent collapse, which favors the competitive
accretion model \citep{Pillai11}. On the other hand, Very Large Array
(VLA) observations of IRAS 05345+3157 \citep{Fontani12} suggest that
turbulence is an important factor in the initial fragmentation of the
parental clump and is strong enough to provide support against further
fragmentation down to thermal Jeans masses.

W3~IRS5 is a massive star-forming region located in the Perseus arm at
a distance of $1.83\pm0.14$ kpc \citep{Imai00} with a total luminosity
of $2\times10^5$ $L_{\sun}$ \citep{Campbell95}. A cluster of
hypercompact ($< 240$ AU) H~II regions are found from high-resolution
cm-wavelength observations
\citep{Claussen94,Wilson03,vanderTak05}. Based on observations at
near-IR wavelengths, \citet{Megeath96,Megeath05,Megeath08} found a high stellar
surface density of $\sim10000$ pc$^{-2}$ and proposed that W3~IRS5 is
a Trapezium cluster \citep{Abt00} in the making. A high protostellar number density
exceeding $10^{6}$ protostars pc$^{-3}$ is also concluded by
\citet{Rodon08} from their 1.4~mm observations with the Plateau de
Bure Interferometer (PdBI). The proximity and the dense protostellar
environment make W3~IRS5 an excellent target to study massive star
formation in a highly clustered mode.

(Sub)millimeter observations show that the molecular structure of
W3~IRS5 is physically and chemically complex. Multiple bipolar
outflows with various orientations are reported via different tracers
\citep[e.g.][]{Mitchell92,Ridge01,Gibb07,Rodon08,YWang12}. Although
the outflow driving sources are not determined unambiguously, the
infrared pair NIR1 and NIR2 \citep{Megeath05} are likely
candidates. The detection of near-IR and X-ray emission toward W3~IRS5
\citep{Megeath05,Hofner02} likely benefits from an outflow oriented
near the line of sight. A velocity gradient in the NW--SE direction on
scales of a few arcseconds is seen in SO$_2$, which may indicate
rotation of the common envelope of NIR1 and NIR2
\citep{Rodon08,YWang12}. At $14\arcsec$ resolution, many molecular
species are detected toward W3~IRS5, especially sulfur-bearing
molecules \citep{Helmich94,Helmich97}, implying active hot-core
chemistry \citep[e.g.][]{Charnley97} or shocks
\citep[e.g.][]{Pineau-des-Forets93}. The sulfur-bearing species SO and
SO$_2$ peak near NIR1/NIR2, while complex organic molecules such as
CH$_3$CN and CH$_3$OH peak at offset positions
\citep{Rodon08,YWang12}.

This paper presents Submillimeter Array (SMA) observations of W3~IRS5
at 354~GHz with an angular resolution of $1\arcsec$--$3\arcsec$,
revealing the complex molecular environment of the proto-Trapezium
cluster and tracing the star formation processes in the dense
cluster-forming core. The observations and data calibration are
summarized in Sect. 2. We present the observational results in
Sect. 3, followed by data analysis in Sect. 4. We discuss our findings
in Sect. 5, and conclude this paper in Sect. 6.

\begin{figure*}[t]
   \centering   
   \includegraphics[width=16cm]{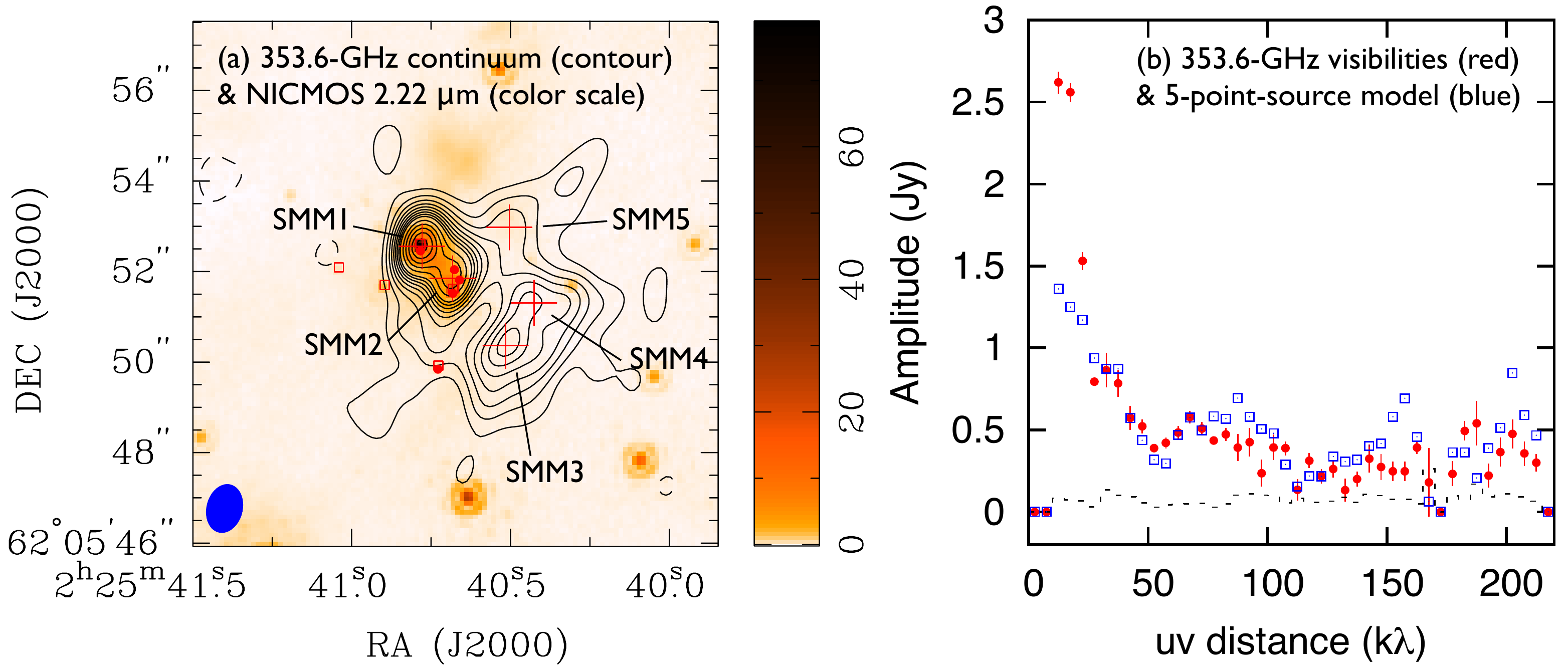}
   \caption{(a) Continuum emission of W3 IRS5 region at 353.6 GHz
     (black contours) overplotted with the NICMOS 2.22 $\mu m$
     emission (color scale). The synthesized beam size is
     $1\farcs1\times0\farcs8$, P.A. $-14\degr$. The contour levels are
     -3, 3, 5, 7, ... , 19, 23, 27, ... $\sigma$ with 1 $\sigma$ of 10
     mJy beam$^{-1}$. The red crosses mark the positions of the five
     point sources derived from the visibility fit. The red open
     squares are the positions of near-infrared sources identified by
     \citet{Megeath05} while the red filled circles represent the
     positions of cm-wave sources from \citet{vanderTak05}. (b)
     Vector-averaged 353.6 GHz continuum visibilities. The red filled
     circles with error bars are the observed data. The expected zero
     amplitude is shown as a dashed histogram. The model consisting of
     five point sources is plotted with open blue squares. }
   \label{fig:cont_map_vis}
\end{figure*}	

\begin{table*}[]
	\caption{Characteristics of the continuum emission in W3 IRS5 at 353.6 GHz}
	\small{
	\begin{tabular}{lcccccccc}
    \hline \hline
     Source & $\alpha(2000)$ & $\delta(2000)$ & $I_\nu$\tablenotemark{a} & $S_\nu$\tablenotemark{b} & $N(\rm H_2)$\tablenotemark{c} & $M(\rm H_2)$\tablenotemark{d} & n(H$_2$)\tablenotemark{e} & Associated cm-wave, \\
& (hh mm ss) & (dd mm ss) & (Jy beam$^{-1}$) & (Jy) & ($\times10^{23}$ cm$^{-2}$) & ($M_{\sun}$) & ($\times10^{7}$ cm$^{-3}$) & IR, or mm source \\
    \hline
    SMM1 & 02 25 40.779 &  +62 05 52.55 & 0.41 & 0.43 & 4.4 & 0.6 & 3.2 & D2\tablenotemark{f},\ Q5\tablenotemark{g},\ K7\tablenotemark{g},\ MIR1\tablenotemark{g},\ NIR1\tablenotemark{h},\\
    & & & & & & & & MM-1\tablenotemark{i},\ SMS1-MM1\tablenotemark{j}\\
    SMM2 & 02 25 40.679 &  +62 05 51.89 & 0.30 & 0.39 & 3.2 & 0.5 & 2.9 & B\tablenotemark{f},\ Q1-3\tablenotemark{g},\ K2-4\tablenotemark{g},\ MIR2\tablenotemark{g},\ NIR2\tablenotemark{h}, \\
    & & & & & & & & MM-2/3\tablenotemark{i},\ MM-5\tablenotemark{i}\\
    SMM3 & 02 25 40.509 &  +62 05 50.41 & 0.16 & 0.24 & 1.7 & 0.3 & 1.8 & SMS1-MM2\tablenotemark{j}\\
    SMM4 & 02 25 40.431 &  +62 05 51.36 & 0.14 & 0.22 & 1.5 & 0.3 & 1.6 & SMS1-MM2\tablenotemark{j}\\    
    SMM5 & 02 25 40.504 &  +62 05 52.99 & 0.09 & 0.13 & 1.0 & 0.2 & 1.0 & \\
    \hline    
  \end{tabular}
  \tablefoot{
		\tablefoottext{a}{Peak intensity measured from the image.} 
        \tablefoottext{b}{Point source flux density measured from the visibility fit.} 
		\tablefoottext{c}{H$_2$ column density. We assume $T_{\rm d} = 150$ K for all submillimeter sources.} 
		\tablefoottext{d}{H$_2$ mass. We assume $T_{\rm d} = 150$ K for all submillimeter sources.} 
		\tablefoottext{e}{H$_2$ volume density. We assume $T_{\rm d} = 150$ K for all submillimeter sources. A spherical source with a diameter of $1\arcsec$ or 1830 AU is assumed.} 
		\tablefoottext{f}{\citet{Claussen94} and \citet{Wilson03}.} 
		\tablefoottext{g}{\citet{vanderTak05}.} 
		\tablefoottext{h}{\citet{Megeath05}.} 
		\tablefoottext{i}{\citet{Rodon08}.} 
		\tablefoottext{j}{\citet{YWang12}.} 
		}
		}
	\label{tab:cont_property}	
\end{table*}


\section{Observations and Data reduction}
W3 IRS5 was observed with the SMA\footnote{The Submillimeter Array is
  a joint project between the Smithsonian Astrophysical Observatory
  and the Academia Sinica Institute of Astronomy and Astrophysics and
  is funded by the Smithsonian Institution and the Academia Sinica.}
\citep[][]{Ho04} on 2008 January 7 in the compact configuration and on
2006 January 15 in the extended configuration. Both observations were
conducted with 7 antennas and were centered on $\alpha(2000)$ =
02$^{\rm h}$25$^{\rm m}$40$\fs$78, $\delta(2000)$ =
62$\degr$05$\arcmin$52$\farcs$50. The weather conditions were good
with $\tau_{\rm 225 GHz}\sim0.1$. The reference $V_{\rm LSR}$ was set
to $-39.0$ km s$^{-1}$. The receiver was tuned to 353.741 GHz in the upper
sideband, covering frequencies from 342.6 to 344.6 GHz and from 352.6
to 354.6 GHz. The correlator was configured to sample each spectral
window by 256 channels, resulting a velocity resolution of $\sim0.35$
km s$^{-1}$ per channel.  For the compact configuration dataset,
3C454.3 was observed as bandpass calibrator. Gain calibration was
performed by frequent observations of two quasars, 0136+478
($\sim16\degr$ away from the source) and 3C84 ($\sim22\degr$ away from
the source). Uranus was adopted as absolute flux calibrator. For the
extended configuration dataset, 3C273, 3C84 and 3C273 were observed,
respectively, as bandpass, gain and flux calibrators. The adopted
total flux density of 3C273 is 9.6 Jy, which is the mean value
reported for five
observations\footnote{http://sma1.sma.hawaii.edu/callist/callist.html}
during 2006 January 13--30. We estimate that the uncertainty in
absolute flux density is about $20\%$. The $uv$-range sampled by the
combined dataset is $10$ k$\lambda$ ($21\arcsec$) to $210$ k$\lambda$
($1\arcsec$).
	
Data reduction was conducted by using the MIR package
\citep{Scoville93} adapted for the SMA, while imaging and analysis
were performed in MIRIAD \citep{Sault95}. To avoid line
contamination, line-free channels were selected to separate the
continuum and line emission in the visibility domain. The continuum
visibilities for each dataset were self-calibrated using the brightest
clean components to enhance the signal-to-noise ratio. Self-calibrated
datasets were combined to image the continuum. To optimize sensitivity
and angular resolution, a robust weighting\footnote{http://www.atnf.csiro.au/computing/software/miriad/userguide/node107.html} parameter of 0.25 was
adopted for the continuum imaging, resulting in a $1\sigma$ rms noise of
10 mJy per $1\arcsec$ beam. Line identification was conducted by using
the compact configuration dataset only with the aid of spectroscopy
databases of JPL\footnote{http://spec.jpl.nasa.gov/} \citep{Pickett98} and
CDMS\footnote{http://www.astro.uni-koeln.de/cdms/} \citep{Muller05}. 
For those lines that were also clearly detected in the extended-configuration 
dataset, we imaged the combined compact+extended dataset.

\section{Observational results}
\subsection{Continuum emission}
Figure \ref{fig:cont_map_vis} shows the continuum image and
visibilities at 353.6 GHz. At a resolution of $1\farcs1\times0\farcs8$
and P.A. $-14\degr$, five major emission peaks, SMM1 to SMM5, were
identified based on their relative intensities and a cutoff of 9
$\sigma$. We note that irregular, extended emission is also observed
around the identified peaks. More emission peaks may be present in the
diffuse surroundings, which requires better sensitivity, angular
resolution, and image fidelity for confirmation. The total flux
density recorded by the SMA is about 2.5 Jy which is about 6\% of the
flux contained in the SCUBA image at 850 $\mu$m \citep{YWang12}. To
derive the positions and flux densities of the emission peaks, we
fitted the data with five point sources in the visibility domain
(Fig. \ref{fig:cont_map_vis} (b)). The properties of the five sources
are summarized in Table \ref{tab:cont_property}. With this simplified
model, about half of the observed flux originates from the unresolved
sources, while the extended emission detected in baselines of 10--25
k$\lambda$ contributes the other half. The fit deviates significantly from the data at long baselines ($>150$ k$\lambda$) and on short baselines ($<20$ k$\lambda$). The former corresponds to fine image detail that we could fit by adding more point sources, but we find that the S/N does not warrant additional ``sources''. The latter indicates that on scales $>8\arcsec$ there is significant emission that our 5-point model does not represent properly. An extended, Gaussian envelope could match these observations. However, on even larger spatial scales our $uv$ sampling misses even larger amounts of emission. Therefore, we do not aim to model this extended emission, but instead refer to the SCUBA 850 $\mu$m emission to trace this component.

Table \ref{tab:cont_property} lists the cm-wave, (sub)mm and IR
sources associated with SMM1--5
\citep{Claussen94,Wilson03,vanderTak05,Megeath05,Rodon08,YWang12}. In
Fig. \ref{fig:cont_map_vis} (a), we see that the emission decreases
between SMM1/SMM2 and SMM3/SMM4/SMM5, dividing the cloud into two
regions with the eastern part containing clusters of IR and cm-wave
emission as well as X-ray emission \citep{Hofner02}, while the western
part shows no IR and cm-wave sources. This dichotomy implies that the
eastern part of the source is more evolved and less embedded than the
western part. We note that the projected positions of SMM1, SMM2 and
SMM3 are aligned nearly in a straight line which could be the result
of fragmentation at the scale of 1\arcsec--2\arcsec ($\sim$ 1800
AU--3600 AU), although this may be a projection effect.

Because SMM1 and SMM2 contain cm-wave sources, the observed
submillimeter flux may contain non-thermal contributions. We estimate
the free-free contribution from VLA observations at 5, 15 and 22.5 GHz
\citep{Tieftrunk97}. For SMM1 and SMM2, the extrapolated free-free
contributions at 353.6 GHz are 8 mJy and 0.4 mJy, respectively, much
smaller than the observed fluxes (SMM1: 0.43 Jy, SMM2: 0.39 Jy). We therefore conclude that the 353.6
GHz emission is dominated by thermal emission from dust and ignore any
contribution of free-free emission toward SMM1 and SMM2.

\begin{figure*}[htbp]
   \centering   
   \includegraphics[width=18cm]{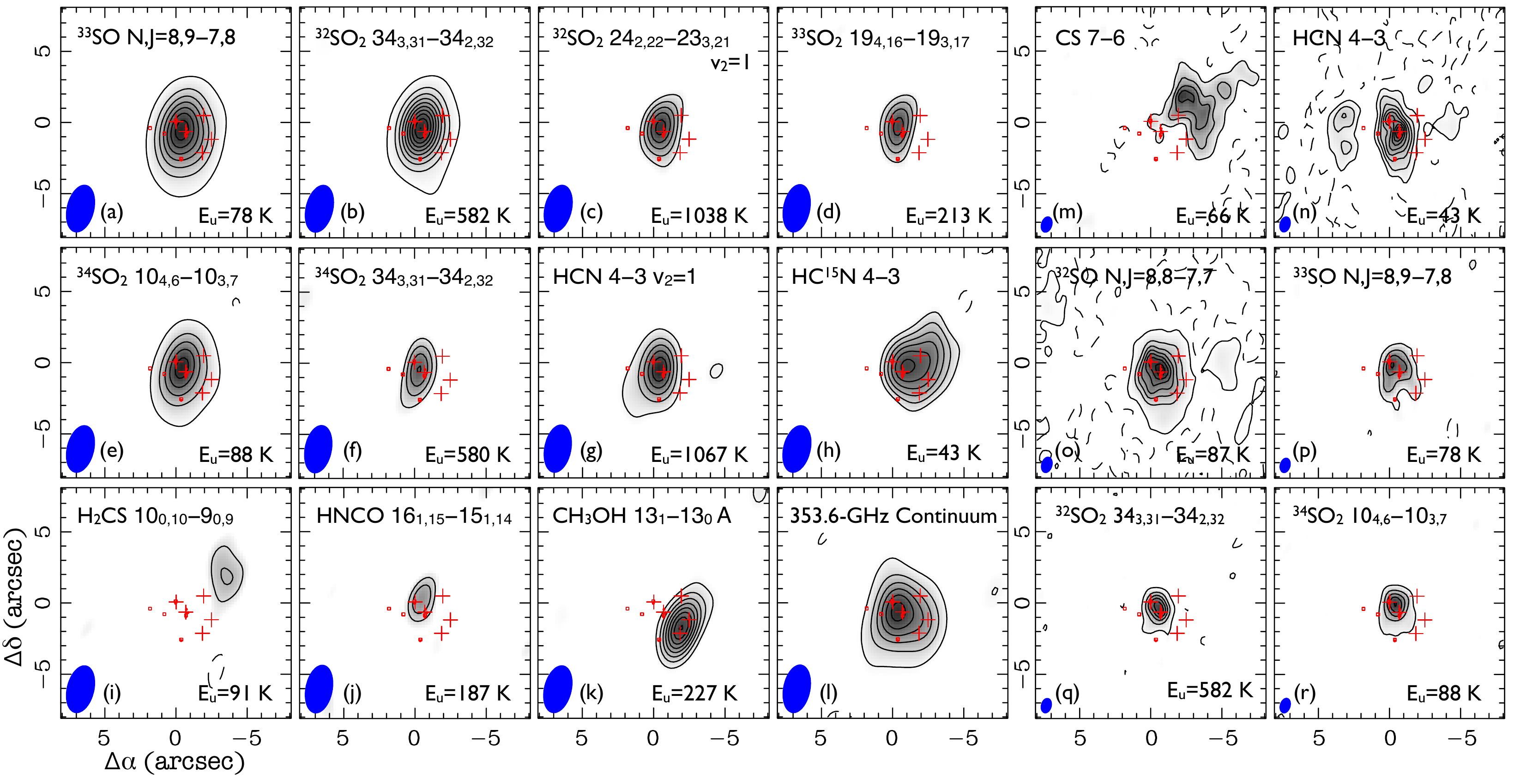}
   \caption{Moment 0 maps of molecular emission toward W3 IRS5. The
     353.6-GHz continuum emission is also presented for comparison. In
     panels (a) to (l), the compact dataset is imaged using natural
     weighting, resulting in a resolution of $3\farcs3\times1\farcs8$,
     (P.A. $-12\degr$) shown as the filled ellipse in the bottom-left
     corner of each panel. In panels (m) to (r), the combined dataset
     is imaged using uniform weighting, resulting in a resolution of
     $1\farcs1\times0\farcs7$ (P.A. $-17\degr$). The five crosses
     represent the positions of SMM1 to SMM5
     (c.f. Fig. \ref{fig:cont_map_vis}). Open squares are the
     positions of IR sources taken from \citet{Megeath05}, while
     filled circles are the positions of cm-wave sources reported by
     \citet{vanderTak05}. The (0, 0) position corresponds to the
     reference phase center (Sect. 2). In all panels, contour levels
     start at 3 $\sigma$ (solid line) and $-$3 $\sigma$ (dashed
     line). The contour units are Jy beam$^{-1}$ km s$^{-1}$ for panel
     (a)--(k) and (m)--(r), and Jy beam$^{-1}$ for panel (l). The
     1-$\sigma$ noise levels from (a) to (r) are 1.0, 0.9, 0.35, 0.39,
     0.72, 0.46, 1.0, 0.45, 0.22, 0.45, 0.36, 0.04, 0.47, 0.94, 0.66,
     0.68, 0.84 and 0.68 respectively. The contour steps from (a) to
     (r) are 10, 10, 2, 2, 10, 2, 4, 2, 1, 1, 2, 4, 4, 5, 20, 5, 6 and
     5, respectively.}
   \label{fig:line_mom0}
\end{figure*}	

\begin{table*}[]

\caption{Detected molecular transitions in W3 IRS5}

\begin{tabular}{lcccc}
\hline \hline
Molecule\tablenotemark{a} & Frequency & Transition & $E_{\rm u}$\tablenotemark{b} & $n_{\rm crit}$\tablenotemark{c} \\
  & (MHz) & & (K) & (cm$^{-3}$) \\
\hline
CS            & 342883.0 & $J$=7--6                                               & 66   & $2.0\times10^7$ 		 \\
SO            & 344310.6 & $N_J$=$8_8$--$7_7$                                       & 87   & $1.2\times10^7$		 \\
$^{33}$SO     & 343086.1 & $N_J$=$8_9$--$7_8$, $F$=15/2--13/2                       & 78   & $1.4\times10^7$		 \\
              & 343087.3 & $N_J$=$8_9$--$7_8$, $F$=17/2--15/2                       & 78  & $1.4\times10^7$ 		 \\
              & 343088.1 & $N_J$=$8_9$--$7_8$, $F$=19/2--17/2                       & 78  & $1.4\times10^7$ 		 \\
              & 343088.3 & $N_J$=$8_9$--$7_8$, $F$=21/2--19/2                       & 78  & $1.4\times10^7$ 		 \\
SO$_2$        & 342761.6 & $J_{K_a,K_c}$=$34_{3,31}$--$34_{2,32}$                 & 582 & \tablenotemark{d} 		 \\          
              & 343923.8 & $J_{K_a,K_c}$=$24_{2,22}$--$23_{3,21}$ $v_2$=1         & 1038 & \tablenotemark{d}		 \\
$^{33}$SO$_2$ & 353741.0 & $J_{K_a,K_c}$=$19_{4,16}$--$19_{3,17}$, $F$=39/2--39/2 & 213  & $2.5\times10^7$		 \\
              & 353741.1 & $J_{K_a,K_c}$=$19_{4,16}$--$19_{3,17}$, $F$=37/2--37/2 & 213  & $2.5\times10^7$ 		 \\
              & 353741.6 & $J_{K_a,K_c}$=$19_{4,16}$--$19_{3,17}$, $F$=41/2--41/2 & 213  & $2.5\times10^7$		 \\
              & 353741.6 & $J_{K_a,K_c}$=$19_{4,16}$--$19_{3,17}$, $F$=35/2--35/2 & 213  & $2.5\times10^7$		 \\
$^{34}$SO$_2$ & 344245.3 & $J_{K_a,K_c}$=$10_{4,6}$--$10_{3,7}$                   & 88   & $3.4\times10^7$		 \\
              & 344581.0 & $J_{K_a,K_c}$=$19_{1,19}$--$18_{0,18}$                 & 167  & $4.8\times10^7$		 \\
              & 353002.4 & $J_{K_a,K_c}$=$14_{7,7}$--$15_{6,10}$                  & 212  & $1.5\times10^7$		 \\
              & 354277.6 & $J_{K_a,K_c}$=$34_{3,31}$--$34_{2,32}$                 & 580  & \tablenotemark{d}		 \\
              & 354397.8 & $J_{K_a,K_c}$=$19_{8,12}$--$20_{7,13}$                 & 326  & $2.9\times10^7$		 \\
HCN           & 354505.5 & $J$=4--3                                               & 43   & $1.8\times10^8$		 \\
              & 354460.5 & $J$=4--3 $v_2$=1                                       & 1067 & \tablenotemark{d}		 \\  
HC$^{15}$N    & 344200.3 & $J$=4--3                                               & 43   & $1.2\times10^8$		 \\ 
H$_2$CS       & 342944.4 & $J_{K_a,K_c}$=$10_{0,10}$--$9_{0,9}$                   & 91   & \tablenotemark{d}		 \\
HNCO          & 352897.9 & $J_{K_a,K_c}$=$16_{1,15}$--$15_{1,14}$, $F$=17--16     & 187  & $3.0\times10^7$		 \\
              & 352897.9 & $J_{K_a,K_c}$=$16_{1,15}$--$15_{1,14}$, $F$=16--15     & 187  & $3.0\times10^7$		 \\
              & 352897.9 & $J_{K_a,K_c}$=$16_{1,15}$--$15_{1,14}$, $F$=15--14     & 187  & $3.0\times10^7$		 \\
CH$_3$OH      & 342729.8 & $J_K$=$13_1$--$13_0$ $A$                               & 227  & $4.8\times10^7$		 \\ 
\hline       
\end{tabular}

\tablefoot{\tablefoottext{a} Spectroscopy data taken from JPL molecular spectroscopy and CDMS. \tablefoottext{b} Upper level energy. \tablefoottext{c} Critical density at 100 K derived from LAMDA \citep{Schoier05}. The critical densities for $^{33}$SO, $^{33}$SO$_2$ and $^{34}$SO$_2$ are quoted from their main isotopologues. \tablefoottext{d} Collisional rate coefficient is not available for this transition.}
\label{tab:line_list}
\end{table*}

\subsection{Line emission}
\subsubsection{Molecular distribution}
Within the passbands (342.6--344.6 GHz and 352.6--354.6 GHz), we
identify 17 spectral features from 11 molecules (Table
\ref{tab:line_list}). Most emission features come from sulfur-bearing
molecules such as CS, SO, $^{33}$SO, SO$_2$, $^{33}$SO$_2$,
$^{34}$SO$_2$ and H$_2$CS. Others come from HCN, HC$^{15}$N, HNCO and
CH$_3$OH. This set of lines covers upper level energies from 43 K to
1067 K and high critical densities ($10^7$--$10^8$ cm$^{-3}$).  Figure
\ref{fig:line_mom0} shows a sample of molecular emission images toward
W3 IRS5 using the compact-configuration data and natural weighting
(panels (a) to (l), with a resolution of $3\farcs3\times1\farcs8$,
P.A. $-12\degr$) and with uniform weighting (panels (m) to (r), with a
resolution of $1\farcs1\times0\farcs7$, P.A. $-17\degr$).
		
\begin{figure}[htbp]
   \centering   
   \includegraphics[width=8cm]{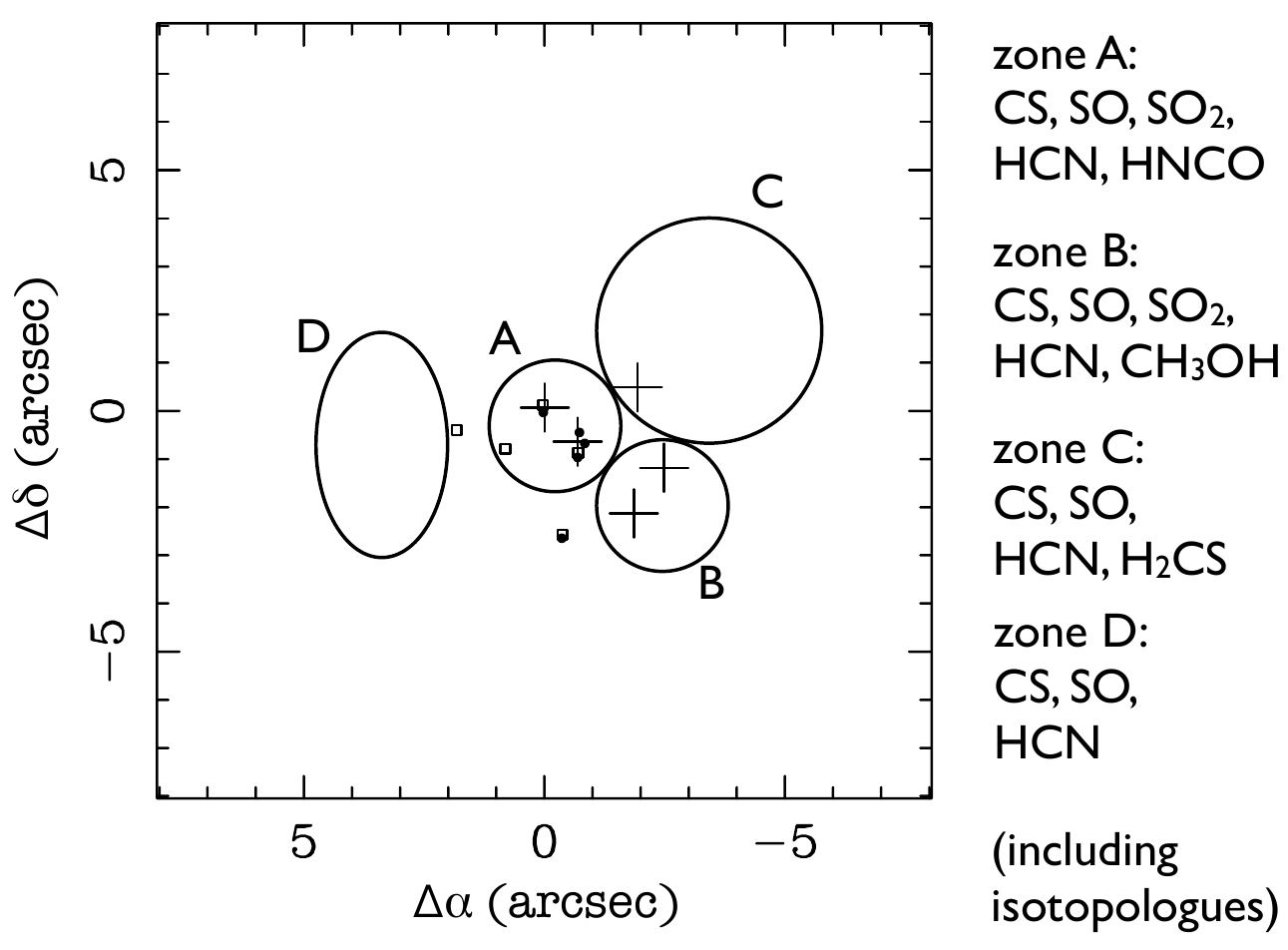}
   \caption{The main molecular zones identified toward W3 IRS5. The
     representative molecular species including their isotopologues
     are listed. The crosses, filled circles and open squares
     represent the same sources as in Fig. \ref{fig:line_mom0}. }
   \label{fig:molecular_zones}
\end{figure}

Most of the emission from SO and SO$_2$ and their isotopologues peaks
toward the continuum peaks SMM1 and SMM2, while weaker emission is
seen toward SMM3 and SMM4. However, other sulfur-bearing molecules
such as CS and H$_2$CS show a very different spatial distribution, and
mainly peak to the north-west of SMM1--5 (Fig. \ref{fig:line_mom0} (i)
and (m)). Most of the emission from HCN and its isotopologues comes
from SMM1 and SMM2 (Fig. \ref{fig:line_mom0} (g), (h) and
(n)). Additional emission can be seen toward SMM3 and SMM4, as well as east
and north-west of the SMM sources. The typical hot-core molecule
CH$_3$OH exclusively peaks toward SMM3 and SMM4
(Fig. \ref{fig:line_mom0} (k)). HNCO is detected toward SMM1 and SMM2
only (Fig. \ref{fig:line_mom0} (j)). Combining all these observational
facts, the overall molecular emission toward W3 IRS5 can be classified
into four zones (A: SMM1/SMM2; B: SMM3/SMM4; C :north-west region; may
be associated with SMM5; and D: east of SMM1--5), and summarized in Figure
\ref{fig:molecular_zones}.

\subsubsection{Velocity channel maps}
\begin{figure*}[htbp]
   \centering   
   \includegraphics[width=17cm]{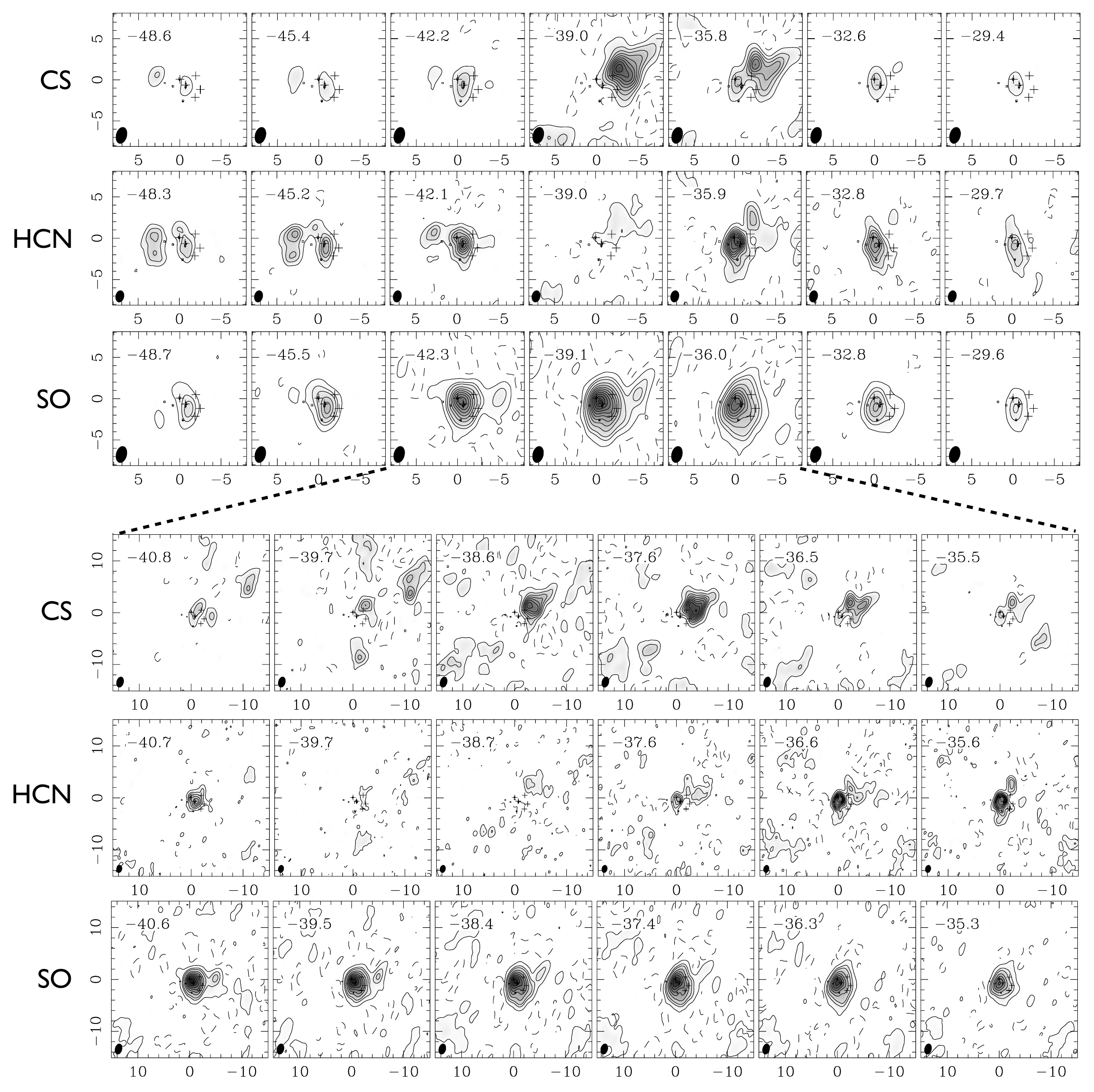}
   \caption{Velocity channel maps of CS $J$=7--6, SO
     $N_J$=$8_8$--$7_7$ and HCN $J$=4--3 toward W3 IRS5. Robust
     weighting (robust = 0) is used for the imaging, resulting
     resolutions of $1\farcs9\times1\farcs2$, P.A. $-15\degr$,
     $1\farcs3\times0\farcs9$, P.A. $-13\degr$, and
     $1\farcs9\times1\farcs2$, P.A. $-14\degr$, for CS, HCN and SO,
     respectively. The $x-$ and $y$-axes are R.A. offset and DEC
     offset relative to the phase center $\alpha(2000)$ =
02$^{\rm h}$25$^{\rm m}$40$\fs$78, $\delta(2000)$ =
62$\degr$05$\arcmin$52$\farcs$50 in arcseconds, respectively. The upper three rows shows the
     velocity range between $-49$ km s$^{-1}$ to $-30$ km
     s$^{-1}$. The lower three rows show a narrower velocity range
     from $-41$ km s$^{-1}$ to $-35$ km s$^{-1}$. The markers are
     identical to Fig. \ref{fig:line_mom0}. Solid and dashed contours
     represent positive and negative intensities, respectively. The
     absolute contour levels for CS in the upper panel are 3, 8,
     13,... $\sigma$ (1 $\sigma$ = 0.11 Jy beam$^{-1}$). The absolute
     contour levels for HCN in the upper panel are 3, 8,
     13,... $\sigma$ (1 $\sigma$ = 0.12 Jy beam$^{-1}$). The absolute
     contour levels for SO in the upper panel are 3, 13, 23, 43,
     63... $\sigma$ (1 $\sigma$ = 0.11 Jy beam$^{-1}$). The absolute
     contour levels for CS in the lower panel are 3, 8,
     13,... $\sigma$ (1 $\sigma$ = 0.18 Jy beam$^{-1}$). The absolute
     contour levels for HCN in the lower panel are 3, 7,
     11,... $\sigma$ (1 $\sigma$ = 0.16 Jy beam$^{-1}$). The absolute
     contour levels for SO in the lower panel are 3, 13, 23, 43,
     63,... $\sigma$ (1 $\sigma$ = 0.14 Jy beam$^{-1}$).}
   \label{fig:CS_HCN_SO_chmap}
\end{figure*}

Among the detected spectral features, most of the emission is compact
(in zones A and B). Only CS $J$=7--6, SO $N_J$=$8_8$--$7_7$ and HCN
$J$=4--3 are extended and trace the large scale gas kinematics.
Figure \ref{fig:CS_HCN_SO_chmap} shows the channel maps of these
molecules from $-49$ km s$^{-1}$ to $-30$ km s$^{-1}$ and from $-41$
km s$^{-1}$ to $-35$ km s$^{-1}$, using robust weighting (robust =
0). These three molecules show emission over a broad velocity
range. The emission at extreme velocities is mainly concentrated
toward zone A. Weaker emission is seen toward zone B. Zone C is best
traced by CS with some emission from SO, HCN, and H$_2$CS as well. At velocities
ranging from $-49$ to $-42$ km s$^{-1}$, the emission from all three
molecules can be seen toward zone D.
		
Of these three molecules, the image quality near the systemic velocity
($\sim-39.0$ km s$^{-1}$) is poor due to the missing flux of extended
emission. However, some additional emission features can be seen near
the systemic velocity. Figure \ref{fig:CS_HCN_SO_chmap} shows the
channel maps of CS, SO and HCN at velocities ranging from $-41$ km
s$^{-1}$ to $-35$ km s$^{-1}$. The emission of HCN is much weaker than
the other species near $-39$ km~s$^{-1}$ suggesting that HCN is much
more extended, and therefore more heavily filtered out by the
interferometer, or that the emission is self-absorbed. Additional
extended emission in the N--S and NW--SE directions can be seen in the
CS channel maps from $-41$ km s$^{-1}$ to $-38$ km s$^{-1}$, which
might delineate two collimated outflows or a single wide-angle outflow
in the NW--SE direction. The small velocity range suggests that the
outflow may be in the plane of sky. In the SO channel maps, a bar-like
structure can be seen from $-41$ km s$^{-1}$ to $-37$ km s$^{-1}$. Our
data suggest that CS, SO and HCN trace different parts of the
molecular condensation in W3 IRS5 and show a complex morphology in
velocity space.

\subsubsection{Line profiles}
\begin{figure}[htbp]
   \centering   
   \includegraphics[width=9cm]{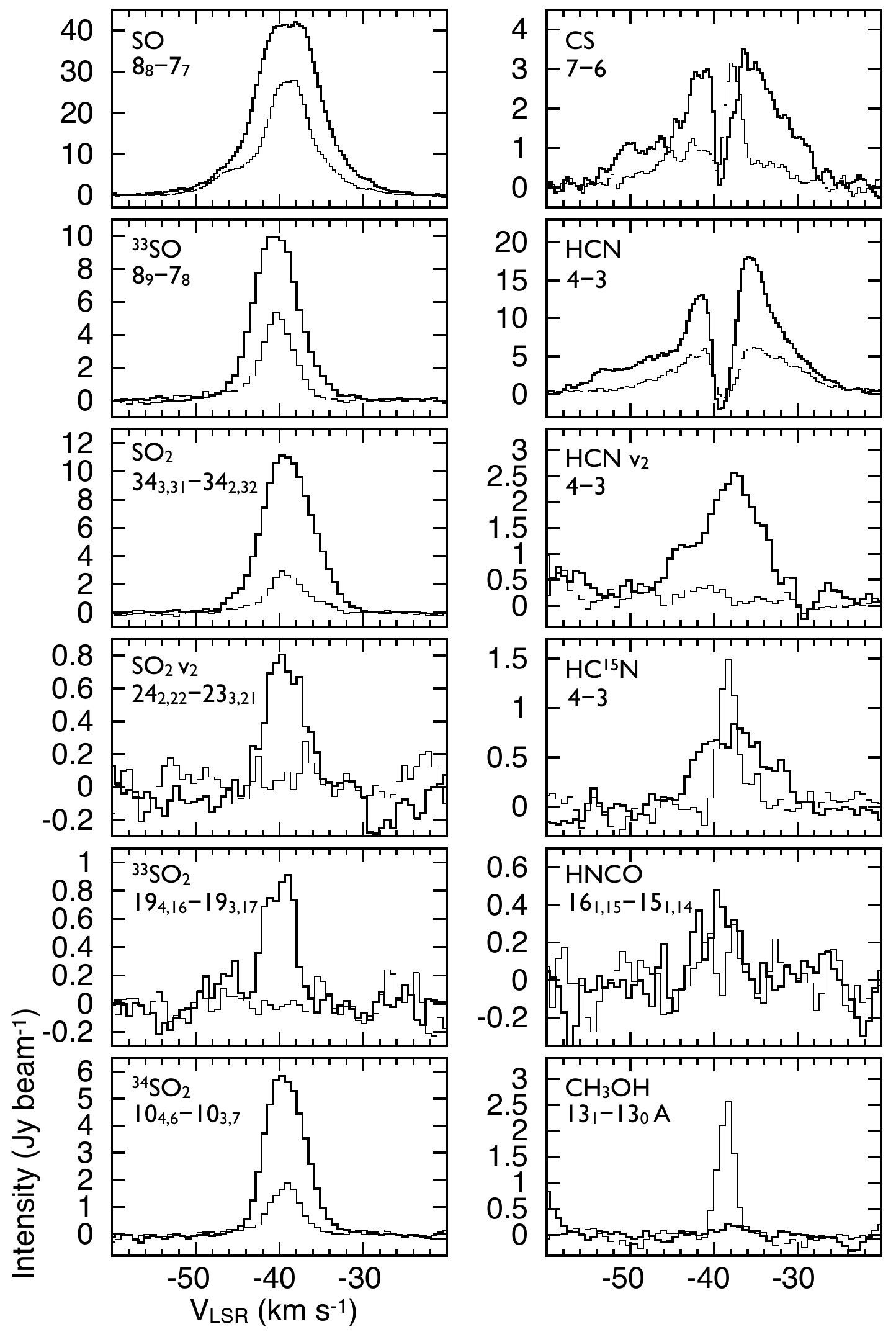}
   \caption{Hanning-smoothed molecular spectra toward molecular zone A
     (thick line; offset position ($-0.42\arcsec$,$-0.43\arcsec$)) and
     B (thin line; offset position
     ($-1.89\arcsec$,$-1.78\arcsec$)). The compact-configuration
     dataset and natural weighting is used for all molecules except CS
     7--6, SO $N_J$=$8_8$--$7_7$ and HCN 4--3 which are taken from the
     combined dataset using natural weighting but convolved to the
     beam size of the compact dataset ($3\farcs3\times1\farcs8$,
     P.A. $-12\degr$). The apparent offset in velocity seen in
     $^{33}$SO and $^{33}$SO$_2$ is due to the blending of hyperfine
     transitions for which the reference velocity is set to the
     transition with lowest frequency (see Table
     \ref{tab:line_list}).}
   \label{fig:mol_hanspec}
\end{figure}

Figure \ref{fig:mol_hanspec} presents the hanning-smoothed spectra
toward zone A (thick line; offset position
$-0.42\arcsec$,$-0.43\arcsec$) and B (thin line; offset position
$-1.89\arcsec$,$-1.78\arcsec$) where most of the detected molecules
peak. The compact-configuration dataset and natural weighting is used
for all molecules except CS, SO and HCN, which are imaged with the
combined dataset in natural weighting and convolved to the same
resolution as the other images.
						
SO shows a flat-top line profile at both positions, implying the line
may be optically thick or is a blend of multiple velocity
components. If SO is fully thermalized, optically thick, and fills the
beam, the peak brightness temperatures indicate kinetic temperatures
of $\sim$70 K and $\sim$45 K for zone A and B, respectively. These
values are lower limits to the kinetic temperature if the beam filling
factor is not unity. In addition, a blue-shifted spectral feature is
seen at zone B.
		
CS and HCN show broad line profiles with a dip near the systemic
velocity due to missing flux of extended emission. Single-dish JCMT
observations of the same transitions \citep{Helmich97} show
single-peaked profiles and rule out self absorption. The line profiles
at zone A are comparable with more emission in the blue-shifted line
wing, suggesting both species trace similar cloud components. At zone B, both lines also show broad line wings. Near systemic velocity, HCN shows a wider dip than CS, suggesting that the spatial distributions of both lines are different. Vibrationally
excited HCN peaks mainly toward zone A and shows an asymmetric line
profile. HC$^{15}$N is present toward both zones; its smaller line
width toward zone B implying that this zone is more quiescent than
zone A.
			
The line profiles of all the other molecules presented in
Fig. \ref{fig:mol_hanspec} are Gaussian toward both zones. There is no
significant velocity shift between the two zones. We note that the
apparent velocity shift seen in $^{33}$SO and $^{33}$SO$_2$ is due to
the blending of multiple hyperfine transitions; we have defined the
velocity axis with respect to the hyperfine component with the lowest
frequency (Table \ref{tab:line_list}).
		
We derive the line center LSR velocity ($V_{\rm LSR}$), FWHM line
width ($dV$) and integrated line intensity ($W$) of the line profiles
in zones A and B using Gaussian decomposition (Table
\ref{tab:line_gaussian}). Since the CS and HCN line profiles are
highly non-Gaussian, we do not attempt to derive the line
parameters. For molecules with hyperfine transitions such as
$^{33}$SO, $^{33}$SO$_2$ and HNCO, we assume that each component has
the same FWHM line width and LSR velocity. The relative intensity of
hyperfine components are calculated assuming LTE and we fit the
combined profile to the data. The results are summarized in Table
\ref{tab:line_gaussian}. From the Gaussian fit, we do not find significant
differences in LSR velocity between zones A and B. Different line
widths are derived in both zones depending on the molecular
transitions but the general trend is that the lines are broader in
zone A, suggesting a more turbulent environment.

We estimate the amount of missing flux of CS 7--6, SO $8_8$--$7_7$, HCN 4--3, and $^{32}$SO$_2$ $34_{3,31}$--$34_{2,32}$ by comparing our SMA observations with the JCMT observations \citep{Helmich97}. Our SMA observations miss $\sim70\%$, $\sim40\%$, and $\sim80\%$ flux for CS, SO, and HCN, respectively. Given the excitation of these lines, the derived missing flux is consistent with the large amount missing flux at 850 $\mu$m ($\sim94\%$; Sect. 3.1). The SMA observation of $^{32}$SO$_2$ $34_{3,31}$--$34_{2,32}$ likely recovers all emission, because this transition has a high upper level energy (corresponding to 582 K), and the emission therefore preferentially traces the dense, warm, and compact part of W3 IRS5.

\begin{table*}[]

\caption{Gaussian decomposition of the line profiles in W3 IRS5}
\small
\begin{tabular}{lcccccccc}
\hline \hline
         &        & \multicolumn{3}{c}{Zone A:SMM1/SMM2} & & \multicolumn{3}{c}{Zone B:SMM3/SMM4} \\ 
\cline{3-5} \cline{7-9}
Molecule & Transition & $W$\tablenotemark{a} & $\Delta V$\tablenotemark{b} & $V_{\rm LSR}$\tablenotemark{c} & & $W$\tablenotemark{a} & $\Delta V$\tablenotemark{b} & $V_{\rm LSR}$\tablenotemark{c}\\
         &     & (Jy beam$^{-1}$ km s$^{-1}$) & (km s$^{-1}$) & (km s$^{-1}$) & & (Jy beam$^{-1}$ km s$^{-1}$) & (km s$^{-1}$) & (km s$^{-1}$) \\
\hline
CS\tablenotemark{d}            & 7--6 & \nodata       & \nodata     & \nodata       & & \nodata       & \nodata     & \nodata       \\
SO     & $8_8$--$7_7$ & $412.9\pm2.2$ & $9.0\pm0.1$ & $-39.0\pm0.1$ & & $207.8\pm2.5$ & $7.4\pm0.1$ & $-39.0\pm0.1$ \\
$^{33}$SO\tablenotemark{e}     & $8_9$--$7_8$ & $68.9\pm0.5$  & $6.0\pm0.1$ & $-39.1\pm0.1$ & & $25.7\pm0.6$  & $4.4\pm0.1$ & $-38.9\pm0.1$ \\
SO$_2$ & $34_{3,31}$--$34_{2,32}$ & $83.8\pm0.6$  & $7.1\pm0.1$ & $-39.1\pm0.1$ & & $15.6\pm0.5$  & $5.6\pm0.2$ &	$-39.1\pm0.1$ \\          
              & $24_{2,22}$--$23_{3,21}$ $v_2$=1 & $4.4\pm0.3$   & $5.1\pm0.4$ & $-39.4\pm0.2$ & & \nodata       & \nodata     & \nodata       \\
$^{33}$SO$_2$\tablenotemark{e} & $19_{4,16}$--$19_{3,17}$ & $4.1\pm0.3$   & $4.1\pm0.4$ & $-39.8\pm0.2$ & & \nodata       & \nodata     & \nodata       \\
$^{34}$SO$_2$ & $10_{4,6}$--$10_{3,7}$ & $35.1\pm0.4$  & $5.5\pm0.1$ & $-39.3\pm0.1$ & & $8.3\pm0.3$   & $4.3\pm0.2$ &	$-39.3\pm0.1$ \\
              & $19_{1,19}$--$18_{0,18}$ & $63.0\pm0.5$  & $6.1\pm0.1$ & $-39.3\pm0.1$ & & $14.8\pm0.4$  & $4.1\pm0.1$ &	$-39.4\pm0.1$ \\
              & $14_{7,7}$--$15_{6,10}$ & $2.1\pm0.3$   &	 $6.0\pm1.0$ & $-39.2\pm0.4$ & & \nodata       & \nodata     & \nodata       \\
              & $34_{3,31}$--$34_{2,32}$ & $4.4\pm0.4$   & $4.8\pm0.5$ & $-38.3\pm0.2$ & & \nodata       & \nodata     & \nodata       \\
              & $19_{8,12}$--$20_{7,13}$ & $1.2\pm0.2$   & $2.2\pm0.4$ & $-39.2\pm0.2$ & & \nodata       & \nodata     & \nodata       \\
HCN\tablenotemark{d}           & 4--3 & \nodata       & \nodata     & \nodata       & & \nodata       & \nodata     & \nodata\\
              & 4--3 $v_2$=1 & $24.5\pm1.0$  & $10.2\pm0.5$& $-38.2\pm0.2$ & & \nodata       & \nodata     & \nodata       \\  
HC$^{15}$N    & 4--3 & $7.8\pm0.4$   & $9.5\pm0.5$ & $-37.7\pm0.2$ & & $3.7\pm0.2$	 & $2.3\pm0.2$ & $-38.3\pm0.1$ \\ 
H$_2$CS\tablenotemark{f}       & $10_{0,10}$--$9_{0,9}$ & \nodata       & \nodata     & \nodata       & & \nodata       & \nodata     & \nodata        \\
HNCO\tablenotemark{e}          & $16_{1,15}$--$15_{1,14}$ & $2.0\pm0.3$   & $4.9\pm1.0$ & $-39.4\pm0.4$ & & \nodata       & \nodata     & \nodata       \\
CH$_3$OH      & $13_1$--$13_0$ $A$ & $0.7\pm0.2$   & $3.4\pm1.3$ & $-37.7\pm0.5$ & & $6.1\pm0.2$	 & $2.3\pm0.1$ & $-38.5\pm0.1$ \\ 
\hline       
\end{tabular}

\tablefoot{The beam size for all transitions listed in this table is about $3\farcs3\times1\farcs8$, P.A. $-12\degr$. \tablefoottext{a} Integrated intensity. \tablefoottext{b} FWHM line width. \tablefoottext{c} Line center LSR velocity. \tablefoottext{d} Not attempted to perform Gaussian decomposition due to missing flux and complex line profile. \tablefoottext{e} With hyperfine transitions. Total $W$ is reported. \tablefoottext{f} Peaked toward zone C.}
\label{tab:line_gaussian}
\end{table*}

\section{Analysis}
\subsection{Mass and density of the submillimeter sources}
We estimate the H$_2$ column density and mass of each point source
based on the continuum at 353.6 GHz. The H$_2$ column density at the
emission peak can be estimated as
	\begin{equation}
		N({\rm H_2}) = \frac{I_\nu a}{2 m_{\rm H} \Omega_b \kappa_\nu B_\nu(T_{\rm d})},
	\end{equation}
where $I_\nu$ is the peak flux density, $a$ is the gas-to-dust ratio
(100), $\Omega_b$ is the beam solid angle, $m_{\rm H}$ is the mass of
atomic hydrogen, $\kappa_\nu$ is the dust opacity per unit mass and
$B_\nu(T_{\rm d})$ is the Planck function at dust temperature $T_{\rm
  d}$. We apply the interpolated value of $\kappa_\nu(845 {\rm \mu m})
\approx 2.2$ cm$^2$ g$^{-1}$ as suggested by \citet{Ossenkopf94} for
gas densities of $10^6$--$10^8$ cm$^{-3}$ and coagulated dust
particles with thin ice mantles. The dust temperature of all five
submillimeter sources is assumed to be 150 K (see excitation analysis,
Sect. 4.2). The gas mass for each continuum peak is estimated from the
total flux derived from the visibility fit (Table
\ref{tab:cont_property}) via
	\begin{equation} 
		M_{\rm gas} = \frac{S_\nu d^2 a}{\kappa_\nu B_\nu(T_{\rm d})},
	\end{equation}
where $S_\nu$ is the total flux density of dust emission and $d$ (1.83
kpc) is the distance to the source. Assuming a spherical source with
$1\arcsec$ in size, the H$_2$ number density is also derived. We
summarized the results in Table \ref{tab:cont_property}.
		
From the density and mass estimates, we find beam-averaged H$_2$
column densities, masses and volume densities toward SMM1 and SMM2 of
about $3$--$4\times10^{23}$ cm$^{-2}$, 0.5 $M_{\sun}$ and
$3\times10^7$ cm$^{-3}$, respectively. Toward SMM3, SMM4 and SMM5
values smaller by factors 2--3 are found. The derived H$_2$ column
densities and masses are sensitive to the adopted dust opacity and
temperature. If a bare grain model is adopted \citep{Ossenkopf94}, the
derived values are reduced by a factor of $<$3. For dust temperatures
between 100 and 200 K, the derived values change by a few ten\%
only. If the dust temperature is 30 K, the reported H$_2$ column
densities and masses increase by factors of $\sim$5. The excitation
analysis of Sects.\ 4.2.1 and 4.2.2 suggest such low temperatures are
unlikely. The estimated H$_2$ volume density is sensitive to the
assumed source size as size$^{-3}$. For example, the densities toward
SMM1 and SMM2 increase to $2.4\times10^{8}$ cm$^{-3}$ if the source
size decreases to $0{\farcs}5$. Since the submillimeter continuum
peaks are unresolved, we treat the H$_2$ volume densities of Table
\ref{tab:cont_property} as lower limit. We note that the core masses are all small ($<$1 $M_{\sun}$).

\subsection{Excitation analysis: temperatures}		
The detected molecular transitions, covering $E_{\rm u}$ from 43 K to
1067 K and critical densities of $\sim10^7$--$10^8$ cm$^{-3}$ (Table
\ref{tab:line_list}), form an useful dataset for the diagnostics of
the physical conditions toward W3 IRS5. Among these molecular lines,
multiple detections of the transitions from SO$_2$ and $^{34}$SO$_2$
toward zones A and B allow us to perform excitation analysis. Multiple
detections of CH$_3$CN $J$=12--11 transitions reported by
\citet{YWang12} form another useful dataset for additional constraints
of the excitation conditions toward zone B. We describe the details in
the following two sections.

\begin{figure}[htbp]
   \centering   
   \includegraphics[width=8cm]{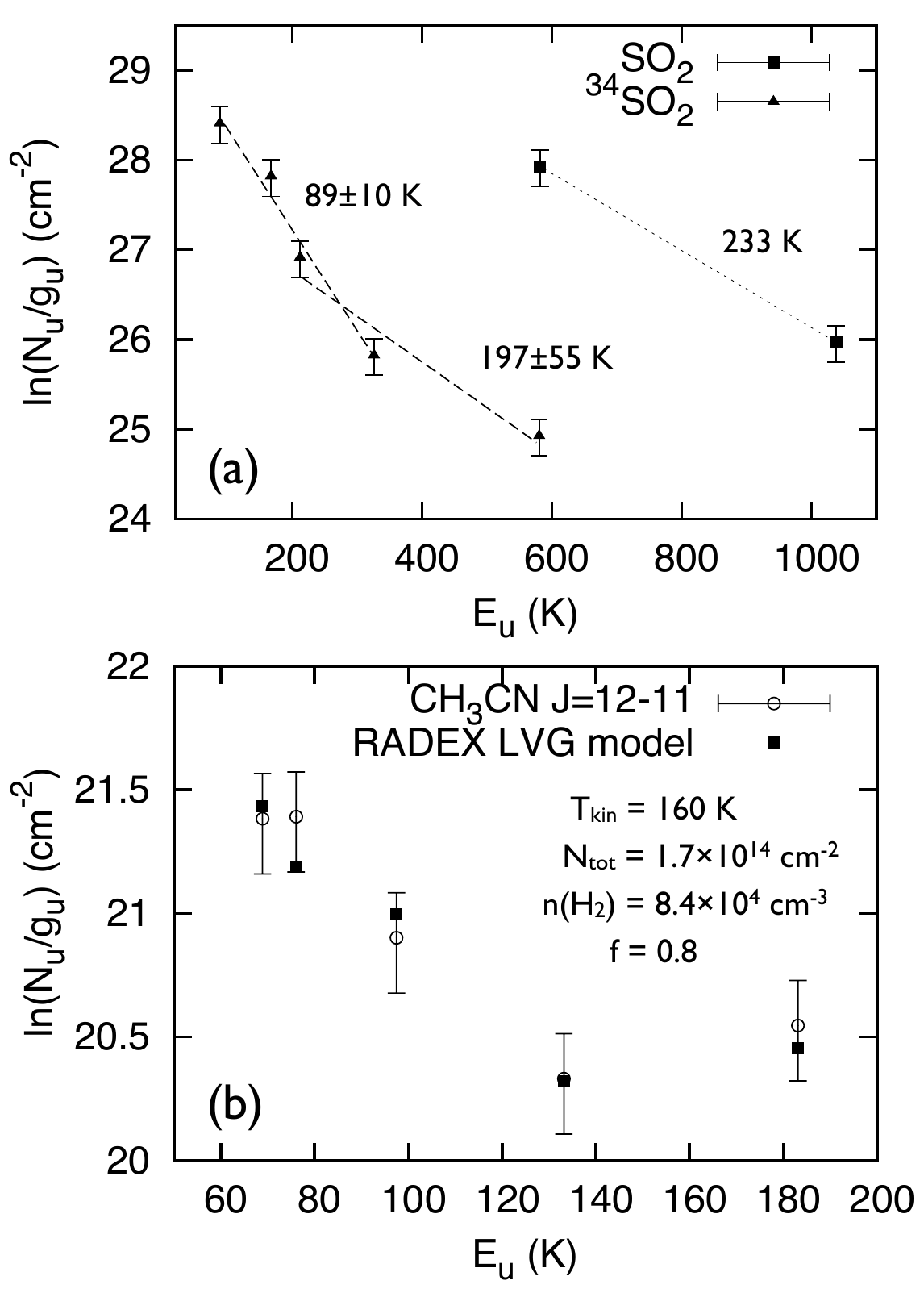}
   \caption{(a) Rotation diagram of SO$_2$ (filled squares) and
     $^{34}$SO$_2$ (filled triangles) toward zone A. The straight
     lines are the model fit. The derived rotational temperatures are
     indicated. (b) Rotation diagram of CH$_3$CN $J$=12--11 toward zone B (open
     circles) taken from \citet{YWang12}. The best-fit model from the
     RADEX LVG analysis is overplotted for comparison (filled
     symbols).}
   \label{fig:RD_PD_plot}
\end{figure}	

\subsubsection{Rotational diagram}
The molecular excitation conditions can be estimated via rotation
diagram analysis assuming all the transitions are optically thin and
the emission fills the beam. In a rotation diagram, the column density
of the upper state $N_u$ is given by
\begin{equation}
		\ln(\frac{N_{\rm u}}{g_{\rm u}}) = \ln(\frac{N_{\rm tot}}{Q}) - \frac{E_{\rm u}}{kT_{\rm rot}}, \label{RD_eqn}
\end{equation}			
where $g_{\rm u}$ is the total degeneracy of the upper state, $N_{\rm
  tot}$ is the total molecular column density, $Q$ is the partition
function, $E_{\rm u}$ is the upper level energy, $k$ is the Boltzmann
constant and $T_{\rm rot}$ is the rotational temperature. The
left-hand side of equation \ref{RD_eqn} can be derived from the
observations via
\begin{equation}
		\ln(\frac{N_{\rm u}^{\rm obs}}{g_{\rm u}}) = \ln(\frac{2.04\times10^{20}}{\theta_a\theta_b}\frac{W}{g_Ig_K\nu_0^3S\mu_0^2}) {\ \rm cm^{-2}}, \label{RD_obs_eqn}
\end{equation}	
where $\theta_a$ and $\theta_b$ are the major and minor axes of the
clean beam in arcsec, respectively, $W$ is the integrated intensity in
Jy beam$^{-1}$ km s$^{-1}$, $g_I$ and $g_K$ are the spin and projected
rotational degeneracies, respectively, $\nu_0$ is the rest frequency
in GHz, $S$ is the line strength and $\mu_0$ is the dipole moment of
the transition in Debye. Figure \ref{fig:RD_PD_plot} plots the
logarithm of the column densities from our data (Eq. \ref{RD_obs_eqn})
versus $E_{\rm u}/k$ (Eq. \ref{RD_eqn}) and a fitted straight line with
$T_{\rm rot}$ and $N_{\rm tot}$ as free parameters. We adopt a $20\%$
uncertainty of the integrated intensity in the analysis.

\begin{figure*}[htbp]
   \centering   
   \includegraphics[width=16cm]{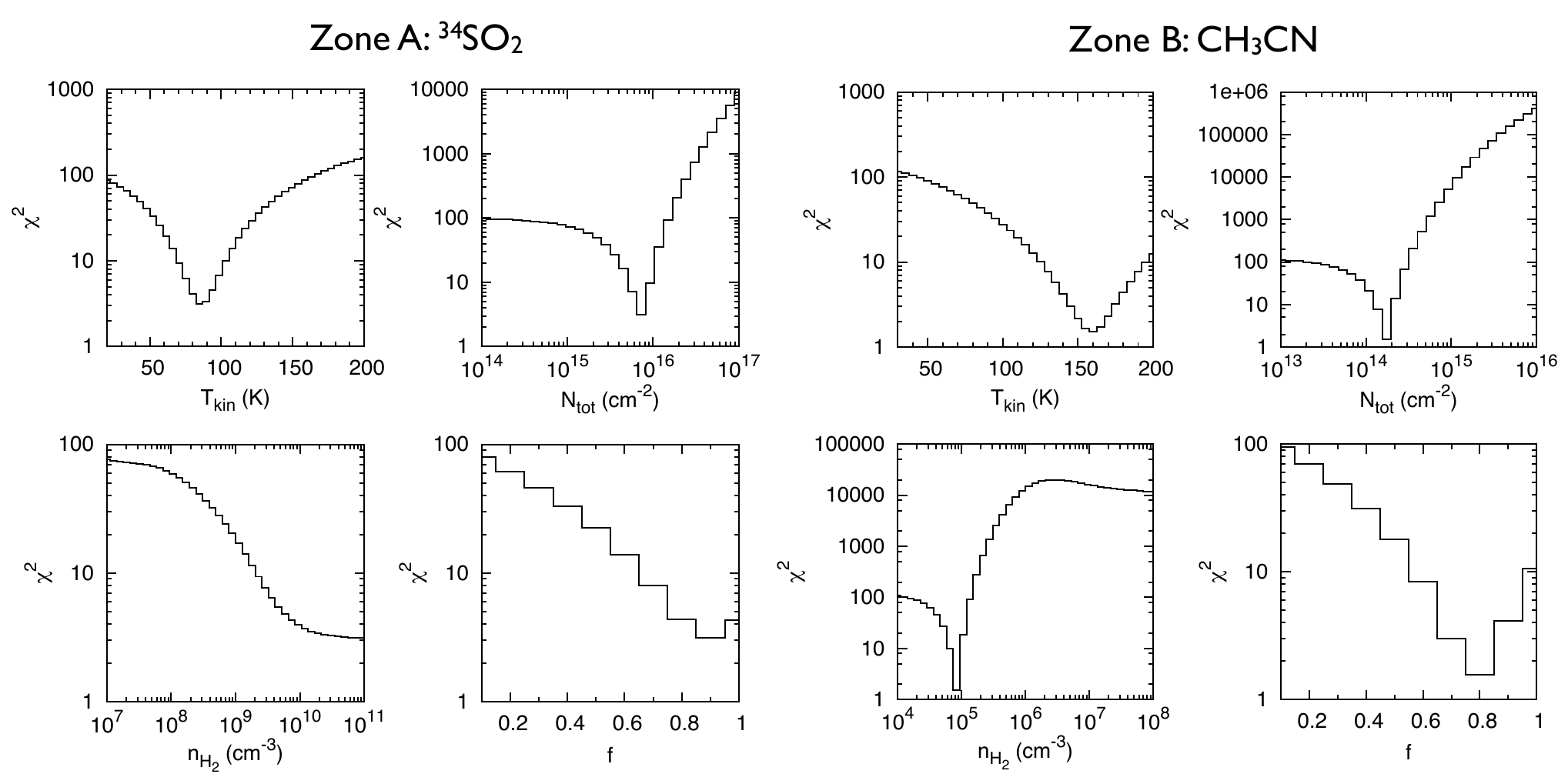}
   \caption{RADEX excitation analysis toward zone A traced by
     $^{34}$SO$_2$ (left four panels) and zone B traced by CH$_3$CN
     (right four panels). The $\chi^2$ surfaces near the best-fit
     solution on each parameter axis are displayed. }
   \label{fig:RADEX_results}
\end{figure*}	

Figure \ref{fig:RD_PD_plot} (a) shows the rotation diagrams of SO$_2$
and $^{34}$SO$_2$ toward zone A, which sample level energies from 88 K
to 1038 K (Table \ref{tab:line_list}). $^{34}$SO$_2$ is detected in
five transitions with $E_{\rm u}$ ranging from 88 to 580 K, and
indicates a $T_{\rm rot}$ of $141\pm25$ K. However, the curvature in
the distribution of the points implies that there might be a cooler
and a warmer component along the line of sight. A two-component fit to
the data gives $T_{\rm rot}=89\pm 10$ K and $T_{\rm rot}=197\pm 55$
K. The gas component traced by SO$_2$ ($E_{\rm u}$ = 582 and 1038 K)
implies a rotational temperature of 233 K, compatible with the
temperature of the warm component derived from $^{34}$SO$_2$ but
obviously less well constrained because only two data points are
available.

The curvature in the $^{34}$SO$_2$ rotation diagram
(Fig. \ref{fig:RD_PD_plot} (a)) may also be due to line opacity (unlikely
for the $^{34}$SO$_2$ isotopologue) or subthermal excitation (see next
section). The relatively abundant main-isotope SO$_2$ may have
optically thick lines, leading to an overestimate for $T_{\rm
  rot}$. However, we detected the $34_{3,31}-34_{2,32}$ transition of
both SO$_2$ and $^{34}$SO$_2$. Equation \ref{RD_obs_eqn} yields a
$N_{\rm u}$(SO$_2$)/$N_{\rm u}$($^{34}$SO$_2$) ratio of $\sim20$ at
$50-300$ K, consistent with the $^{32}$S/$^{34}$S ratio of 22 in the
interstellar medium \citep{Wilson94}, suggesting that the
$34_{3,31}-34_{2,32}$ transition of SO$_2$ (and its isotopologues) is
optically thin. The $24_{2,22}-23_{3,21} v_2=1$ transition of SO$_2$
is also likely optically thin since its Einstein-A coefficient is only
about a factor of 2 greater than that of the $34_{3,31}-34_{2,32}$
transition. These considerations suggest the a temperature gradient in
zone A explains the rotation diagrams, with a cooler region at
$\sim100 K$ and a warmer region of $\ge200 K$ (Table
\ref{tab:excitation}).

We do not have enough data to obtain a reliable temperature estimate
toward zone B (Table \ref{tab:line_gaussian}). Only the lowest two
transitions of $^{34}$SO$_2$ in level energy are detected here, and
imply a rotational temperature of about 133 K. To have a better
constraint of the excitation conditions toward zone B, we took the
CH$_3$CN $J=12-11$ lines obtained by \citet{YWang12} with the SMA at a
resolution of $4.0\arcsec\times2.6\arcsec$. This molecule peaks toward
zone B exclusively and shows a much narrower FWHM line width
($\sim1.6$ km s$^{-1}$) compared to $^{34}$SO$_2$ ($\sim4$ km
s$^{-1}$), suggesting a different molecular condensation in zone B.
Figure \ref{fig:RD_PD_plot} (b) shows the CH$_3$CN $J=12-11$ $K$=0--4
rotation diagram. The scatter in the data indicates that the lines are
not optically thin. Therefore, we can not carry out rotation diagram
analysis on the CH$_3$CN lines, and instead perform an LVG analysis in
Sect. 4.2.2.

\subsubsection{RADEX LVG model}
To further investigate the physical conditions such as kinetic
temperature ($T_{\rm kin}$), column density ($N_{\rm tot}$) and H$_2$
volume density ($n_{\rm H_2}$) toward zone A and B, we performed
statistical equilibrium calculations to solve the level populations
inferred from the observations. The purpose here is to have a crude estimate. Detailed analysis of temperature and density profiles in W3 IRS5 is beyond the scope of this work. We used the code RADEX
\citep{vanderTak07} which solves the level populations with an escape
probability method in the large-velocity-gradient (LVG)
regime. Limited by the available observed data and collisional rate
coefficients \citep{Schoier05}, we perform calculations only for the
$^{34}$SO$_2$ lines and CH$_3$CN $J=12-11$ lines \citep{YWang12}. For
$^{34}$SO$_2$, we adopted the collisional rate coefficients from
SO$_2$ \citep{Green95}\footnote{Recent calculations by \citet{Spielfiedel09}   and \citet{Cernicharo11} show that the collisional rate coefficients calculated by \citet{Green95} are lower by a factor 3--5. The listed critical densities in Table \ref{tab:line_list} may be 3--5 times smaller.} as an approximation since both species have
similar molecular properties. The collisional rate coefficients of
CH$_3$CN are taken from \citet{Green86}. To find the best-fit
solution, we applied $\chi^2$ minimization to the 4-dimensional
parameter space ($T_{\rm kin}$, $N_{\rm tot}$, $n_{\rm H_2}$ and $f$,
the beam filling factor).
			
Figure \ref{fig:RADEX_results} shows the $\chi^2$ surfaces near the
best-fit solution on each parameter axis, and Table
\ref{tab:excitation} summarizes the results. Toward zone A, we use
four $^{34}$SO$_2$ lines (excluding the transition with $E_{\rm u} =
580$ K) in the calculations due to the limited table of collisional
rate coefficients. The RADEX calculations suggest that the kinetic
temperature traced by the $^{34}$SO$_2$ lines is about 90 K,
consistent with the temperature derived from rotation diagram analysis
and suggesting the excitation is thermalized. This requires a high
H$_2$ volume density ($>10^{9-10}$ cm$^{-3}$) toward zone
A. Apparently, the cool component toward zone A is very dense. Toward
zone B, our RADEX calculations indicate a kinetic temperature of
160 K. Interestingly, the H$_2$ volume density traced by CH$_3$CN is
only about 10$^{5}$ cm$^{-3}$, lower than the critical density of a
few $10^{6}$ cm$^{-3}$, implying subthermal excitation. The beam
filling factor is about 0.8.  The best-fit model for CH$_3$CN is
plotted as filled squares in Fig. \ref{fig:RD_PD_plot} (b).
			
\begin{figure}[htbp]
   \centering   
   \includegraphics[width=9cm]{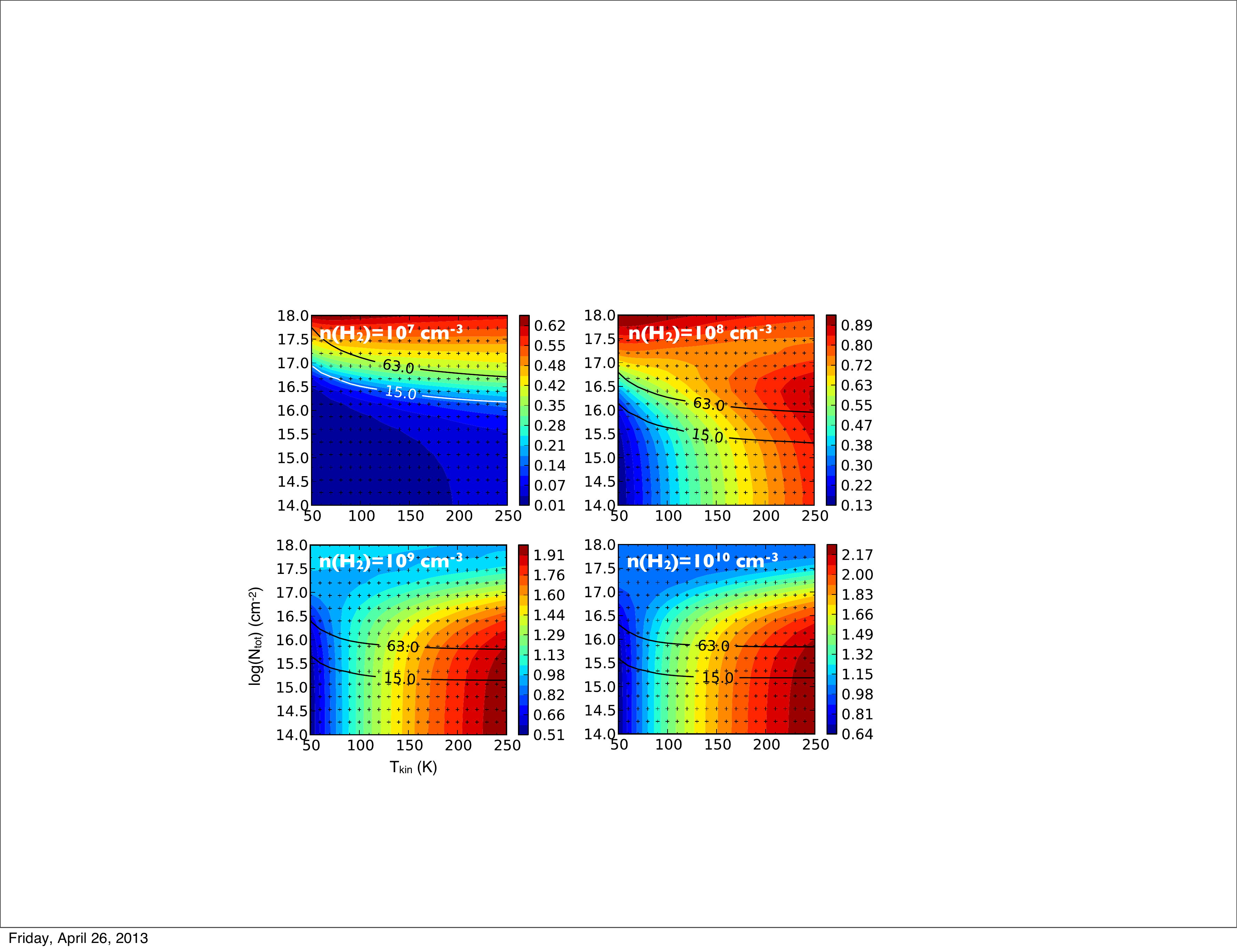}
   \caption{RADEX analysis of the $^{34}$SO$_2$ line ratios toward
     zone A and B assuming a FWHM line width 5.0 km s$^{-1}$. The
     intensity ratios $19_{1,19}-18_{0,18}$/$10_{4,6}-10_{3,7}$ are
     plotted in color scale. The observed integrated intensities of
     the $19_{1,19}-18_{0,18}$ transition are plotted in curves (zone
     A: $\sim63$ Jy beam$^{-1}$ km s$^{-1}$, zone B: $\sim15$ Jy
     beam$^{-1}$ km s$^{-1}$). Toward both zones, the line ratio is
     about 1.8 (Table \ref{tab:line_gaussian}). Our results suggest
     that a high H$_2$ volume density ($>10^{9}$ cm$^{-3}$) is needed
     to reproduce the observed line ratios.}
   \label{fig:34SO2_radex_ratio}
\end{figure}

Alternatively, the excitation conditions toward zone A and B can be
estimated via the line ratio of the $19_{1,19}-18_{0,18}$ over the
$10_{4,6}-10_{3,7}$ transition of $^{34}$SO$_2$ of 1.8. We assume that both transitions have the same filling factor. Using a FWHM
line width of 5.0 km s$^{-1}$ (the average FWHM line widths in zones A
and B are 6.0 km s$^{-1}$ and 4.0 km s$^{-1}$, respectively) in the
RADEX calculations, we calculate the intensity ratio as function of
$T_{\rm kin}$ and $N_{\rm tot}$ for four different values of $n_{\rm
  H_2}$ (Fig. \ref{fig:34SO2_radex_ratio}). We again assume that the two transitions in the RADEX calculations have the same filling factor. These calculations suggest
a high density of $>10^{9}$ cm$^{-3}$ traced by $^{34}$SO$_2$ toward
zone A and B. The kinetic temperatures are $\ge150$ K, depending on
the H$_2$ volume density.
			
Combining the results from the rotation diagram analysis and the RADEX
LVG calculations, we conclude that the gas traced by $^{34}$SO$_2$ is
dense ($n_{\rm H_2}\ge10^{9}$ cm$^{-3}$) in zones A and B. There is a
temperature gradient along the line of sight in zone A, characterized
by a cool region of $\sim100$ K and a warm region of $\ge200$
K. Toward zone B, the temperature gradient is less prominent. However,
a quiescent (FWHM line width 1.6 km s$^{-1}$) and less dense region
($n_{\rm H_2}\sim10^{5}$ cm$^{-3}$) traced by CH$_3$CN is also present in zone
B.

\begin{table*}[]

\caption{Molecular excitations in W3 IRS5}
\small
\begin{tabular}{lcccccccc}
\hline \hline
         &   \multicolumn{3}{c}{Zone A:SMM1/SMM2} & & \multicolumn{3}{c}{Zone B:SMM3/SMM4} \\ 
\cline{2-4} \cline{6-8}
Molecule & $T$ (K) & $N_{\rm tot}$ (cm$^{-2}$) & $X$\tablenotemark{a} & & $T$ (K) & $N_{\rm tot}$ (cm$^{-2}$) & $X$\tablenotemark{b}  & Note \\
\hline
\multicolumn{9}{c}{Rotation diagram: $T$ = $T_{\rm rot}$} \\
\hline
SO$_2$ & 233 & $6.6\times10^{16}$ & $4.4\times10^{-7}$ &  & \nodata & \nodata & \nodata & zone A: 2 trans.\\
$^{34}$SO$_2$ & $89\pm10$ & $6.1^{+3.2}_{-2.2}\times10^{15}$ & $4.1^{+2.1}_{-1.5}\times10^{-8}$ &  & 133 & $1.8\times10^{15}$ & $2.0\times10^{-8}$ & zone A: lowest 4 trans, zone B: 2 trans.\\
              & $197\pm55$ & $3.7^{+5.9}_{-2.4}\times10^{15}$ & $2.5^{+3.9}_{-1.6}\times10^{-8}$ &  & \nodata & \nodata & \nodata & zone A: highest 3 trans.\\
\hline
\multicolumn{9}{c}{RADEX LVG model: $T$ = $T_{\rm kin}$} \\
\hline 
$^{34}$SO$_2$ & 85 & $7.3\times10^{15}$ & $4.9\times10^{-8}$ &  & \nodata & \nodata & \nodata  & zone A: $f=0.9$, $n_{\rm H_2}\ge10^{9}$ cm$^{-3}$ \\
CH$_3$CN\tablenotemark{c} & \nodata & \nodata & \nodata & & 160 & $1.7\times10^{14}$ & \nodata & zone B: $f=0.8$, $n_{\rm H_2}=8.4\times10^{4}$ cm$^{-3}$\\
\hline
\multicolumn{9}{c}{Column densities at $T = T_{\rm rot} = 150$ K} \\
\hline
SO & 150 & $>1.3\times10^{16}$ & $>8.7\times10^{-8}$ & & 150 & $>6.6\times10^{15}$ & $>7.4\times10^{-8}$ & opacities ($\sim0.3-0.5$) in both zones\\
$^{33}$SO & 150 & $9.7\times10^{15}$ & $6.5\times10^{-8}$ & & 150 & $3.6\times10^{15}$ & $4.1\times10^{-8}$\\
SO$_2$ & 150 & $1.4\times10^{17}$ & $9.4\times10^{-7}$  & & 150 & $2.5\times10^{16}$ & $2.8\times10^{-7}$\\
$^{33}$SO$_2$ & 150 & $9.5\times10^{14}$ & $6.4\times10^{-9}$ & & 150 & $<9.5\times10^{13}$ & $<1.1\times10^{-9}$\\
$^{34}$SO$_2$ & 150 & $6.1\times10^{15}$ & $4.1\times10^{-8}$ & & 150 & $1.9\times10^{15}$ & $2.1\times10^{-8}$\\
HC$^{15}$N    & 150 & $1.7\times10^{13}$ & $1.1\times10^{-10}$ & & 150 & $8.1\times10^{12}$ & $9.1\times10^{-11}$\\
HNCO & 150 & $1.1\times10^{14}$ & $7.4\times10^{-10}$ & & 150 & $<3.0\times10^{13}$ & $<3.4\times10^{-10}$\\
CH$_3$OH & 150 & $4.9\times10^{14}$ & $3.3\times10^{-9}$ & & 150 & $4.5\times10^{15}$ & $5.1\times10^{-8}$\\
\hline       
\end{tabular}

\tablefoot{\tablefoottext{a} Fractional abundance. $N$(H$_2$)=$1.5\times10^{23}$ cm$^{-2}$ is derived from our data. \tablefoottext{b} Fractional abundance. $N$(H$_2$)=$8.9\times10^{22}$ cm$^{-2}$ is derived from our data. \tablefoottext{c} Data taken from \citet{YWang12}.}
\label{tab:excitation}
\end{table*}

\subsubsection{Molecular column density}	
We estimate the beam averaged ($3\farcs3\times1\farcs8$) molecular
column density of each detected molecule toward zone A and B via
Eq. \ref{thin_column_density} by assuming optically thin emission and
a single rotational temperature of 150 K which is a representative
value in W3 IRS5 (see previous two sections).
\begin{equation}
		N_{\rm tot} = \frac{N_{\rm u}^{\rm obs}}{g_{\rm u}}\times Qe^{E_{\rm u}/kT_{\rm rot}} \label{thin_column_density}
\end{equation}	
We only use the ground vibrational state for the estimates. If
multiple transitions of a given molecule are detected, we adopt the
averaged value. The uncertainty of the exact rotational temperature
results in an error in the column density of a few 10\% (for $T_{\rm
  rot}$ from 100 K to 200 K). As a consistency check, we estimate the
line center opacity of each transition via
\begin{equation}
		\tau = \frac{c^3}{8\pi\nu_0^3}\frac{A_{\rm ul}}{\Delta V}\frac{g_uN_{\rm tot}}{Q}e^{-E_{\rm u}/kT_{\rm rot}}(e^{h\nu_0/T_{\rm rot}}-1), \label{PD_tau}
\end{equation}
where $c$ is the speed of light, $A_{\rm ul}$ the Einstein-A
coefficient, and $\Delta V$ the FWHM line width. All lines are
consistent with the optically thin assumption ($\tau \ll 1$); only SO
has line center opacities as large as $\tau\sim0.3$--0.5). Therefore,
a lower limit of SO column density is derived. A more representative
value for SO can be derived from $^{33}$SO by assuming the isotopic
ratio $^{32}$S/$^{33}$S of $\sim 132$ \citep{Wilson94,Chin96}. To
convert the molecular column densities into fractional abundances with
respect to H$_2$, we adopt H$_2$ column densities of
$1.5\times10^{23}$ cm$^{-2}$ and $8.9\times10^{22}$ cm$^{-2}$ toward
zone A and B, respectively (derived from the continuum image in
Fig. \ref{fig:line_mom0} (l) with the same assumptions described in
Sect. 4.1). We summarize the results in Table
\ref{tab:excitation}. Toward zone A and zone B, SO and SO$_2$ are very
abundant with fractional abundances up to few $10^{-6}$, several
orders of magnitude higher than found in dark clouds \citep[few
  $10^{-9}$,][]{Ohishi92}. This is indicative of active sulfur
chemistry. Among the detected molecules, we do not see significant
differences in fractional abundance between zones A and B, except for
CH$_3$OH which is a factor of 10 more abundant in zone B.

\begin{figure*}[t]
   \centering   
   \includegraphics[width=18cm]{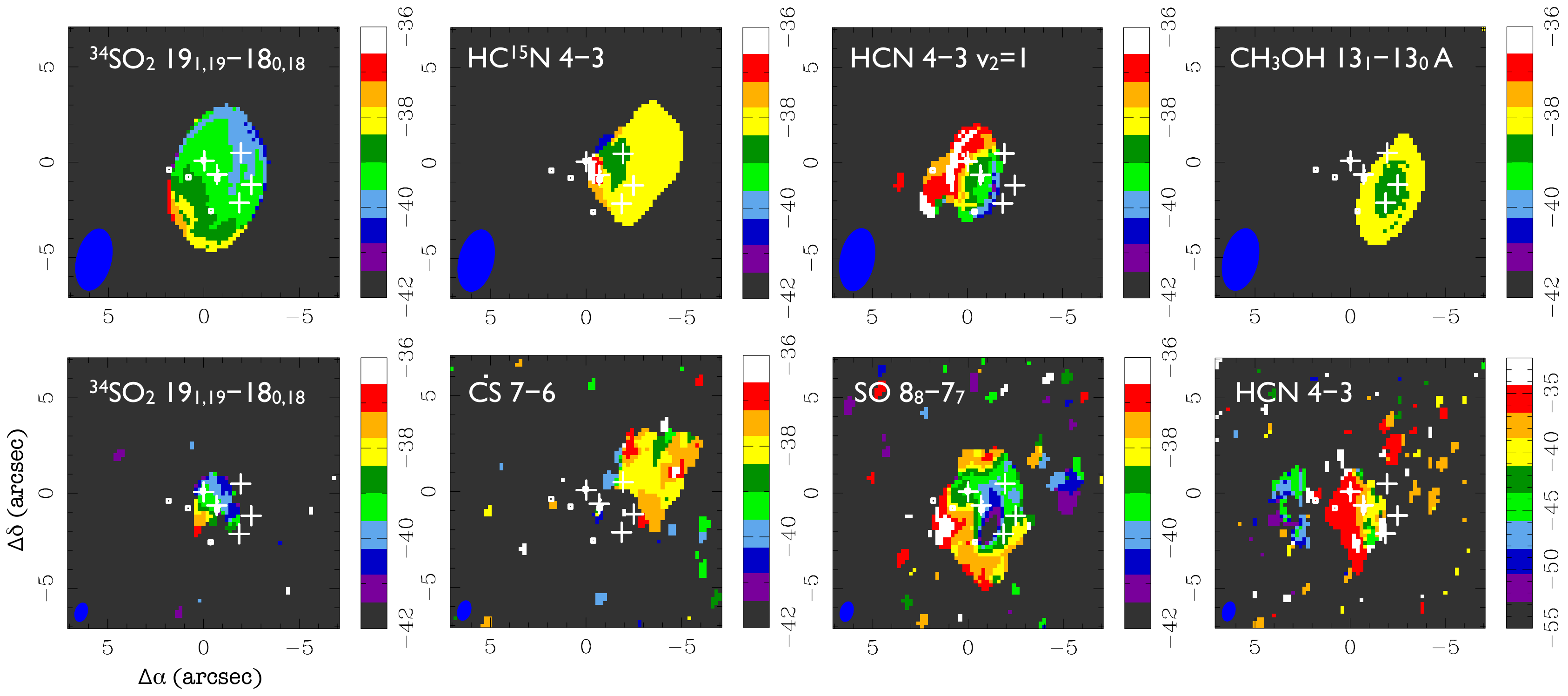}
   \caption{First moment maps of different molecules observed toward
     W3 IRS5. The upper four maps are derived from the
     compact-configuration dataset with natural weighting, while the
     bottom four maps are made from the combined dataset using uniform
     weighting. The markers are identical to the ones in
     Fig. \ref{fig:cont_map_vis}. The zero position is the phase
     center. We note that the velocity ranges are the same for all
     maps except HCN 4--3. }
   \label{fig:line_mom1}
\end{figure*}	

\subsection{Kinematics}
Complex velocity fields in W3 IRS5 are observed in various
molecules. Figure \ref{fig:line_mom1} shows several first moment maps
derived from different molecules. The upper four maps are made from
the compact-configuration dataset using natural weighting, while the
bottom ones are derived from the combined dataset with uniform
weighting. Different velocity gradients are observed in $^{34}$SO$_2$
(NW--SE), HC$^{15}$N (E--W) and HCN $v_2=1$ (NE--SW). CH$_3$OH shows
an interesting velocity distribution with a slightly blue-shifted
emission peak (SMM3 and SMM4). Comparing the velocities near SMM3 and
SMM4 derived from $^{34}$SO$_2$ and CH$_3$OH, we suggest that there
are two distinct regions along the line of sight. Indeed, the line
widths of these two molecules are very different (Table
\ref{tab:line_gaussian}). At one-arcsecond resolution (combined
dataset with uniform weighting), $^{34}$SO$_2$ peaks toward SMM1 and
SMM2, and shows a velocity gradient consistent with the NW--SE
gradient found from lower angular resolution observations. For the
three strongest lines, CS, SO and HCN, we adopted uniform weighting to
form the images in order to minimize the strong sidelobes which
otherwise corrupt the images. CS and HC$^{15}$N show similar velocity
patterns with some patchy velocity shifts. The velocity field traced
by SO is very different from the ones seen from other molecules in
Fig. \ref{fig:line_mom1}, which is likely due to opacity effects. Near
SMM1 and SMM2, HCN and its vibrationally excited transition show similar
velocity patterns. The velocity components traced by CS and HCN toward
zone C are red-shifted with respect to the systemic velocity
($\sim-39$ km s$^{-1}$). However, a clear velocity jump is seen near
SMM5 if we compare the maps of $^{34}$SO$_2$ to CS and HCN, implying
that the gas component in zone C is not closely related to the ones in
zones A and B. Another velocity jump is seen in the HCN map if we
compare the velocities in zones A and D, suggesting that the gas
component in zone D is also not closely related to zones A and B.

\begin{figure}[htbp]
   \centering   
   \includegraphics[width=9cm]{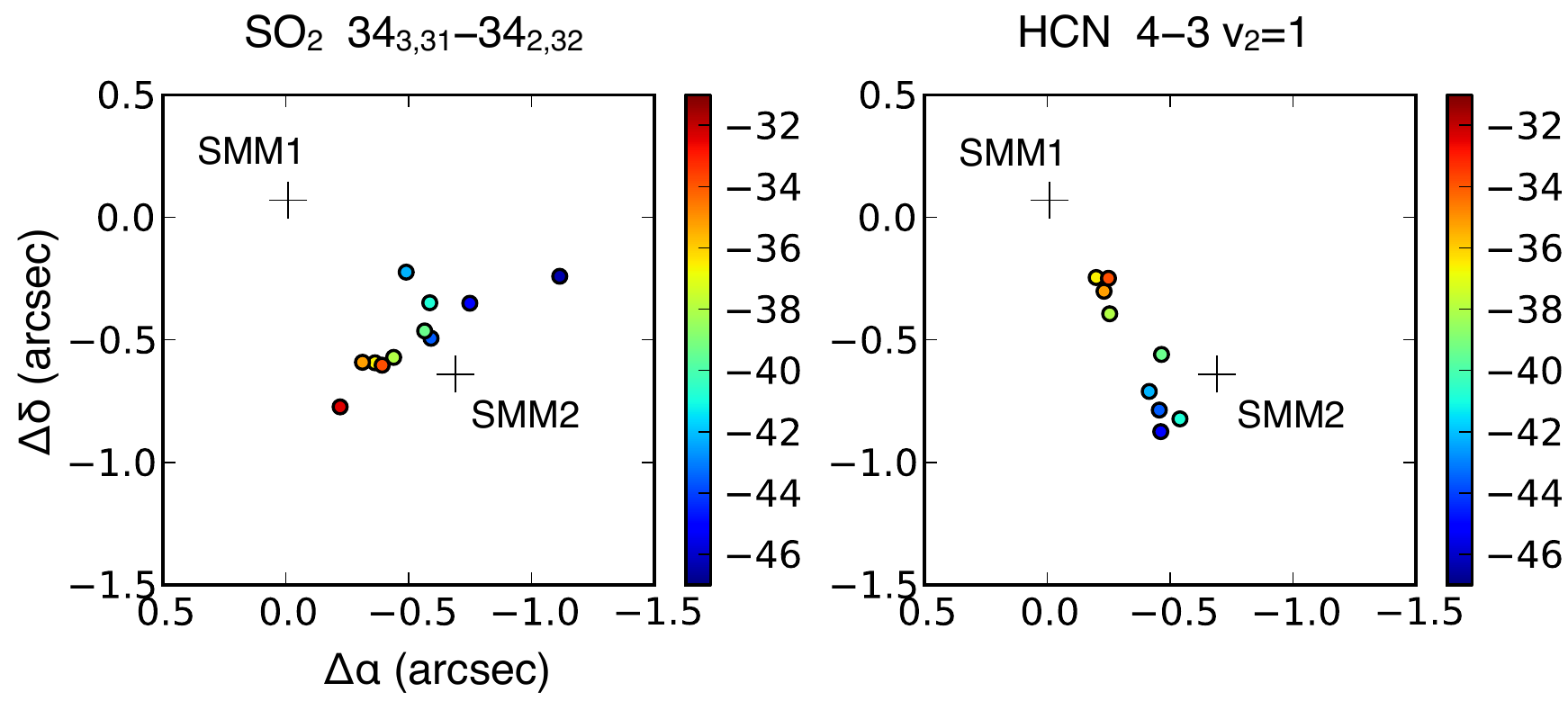}
   \caption{Channel-by-channel locations of the emission peaks
     of the SO$_2$ $34_{3,31}-34_{2,32}$ and HCN 4--3 $v_2=1$. The
     emission peaks are color-coded with respect to their $V_{\rm
       LSR}$ in km s$^{-1}$ shown in the color wedge. Two distinct,
     orthogonal velocity gradients are seen.}
   \label{fig:32SO2_HCNv2_vgrad}
\end{figure}

In Fig. \ref{fig:line_mom1}, we see that the velocity gradients seen
in $^{34}$SO$_2$ and HCN $v_2=1$ are roughly perpendicular to each
other. To confirm this, we fit the emission of HCN 4--3 $v_2=1$
($E_{\rm u}=1067$ K) and SO$_2$ $34_{3,31}-34_{2,32}$ ($E_{\rm u}=582$
K) (compact configuration with natural weighting) channel-by-channel
with a Gaussians to determine the movement of the peak positions. In
Fig. \ref{fig:32SO2_HCNv2_vgrad}, the velocity gradient seen in SO$_2$
is roughly perpendicular to the line joining SMM1 and SMM2, while the
velocity gradient observed in the vibrationally excited HCN is
parallel to this line. The vibrationally excited HCN may trace outflow
or jet emission since the observed velocity gradient is parallel to
the free-free emission knots observed by \citet{Wilson03}. We suggest
that SO$_2$ traces a rotating structure orthogonal to this outflow
axis.
		
\begin{figure}[htbp]
   \centering   
   \includegraphics[width=5.5cm]{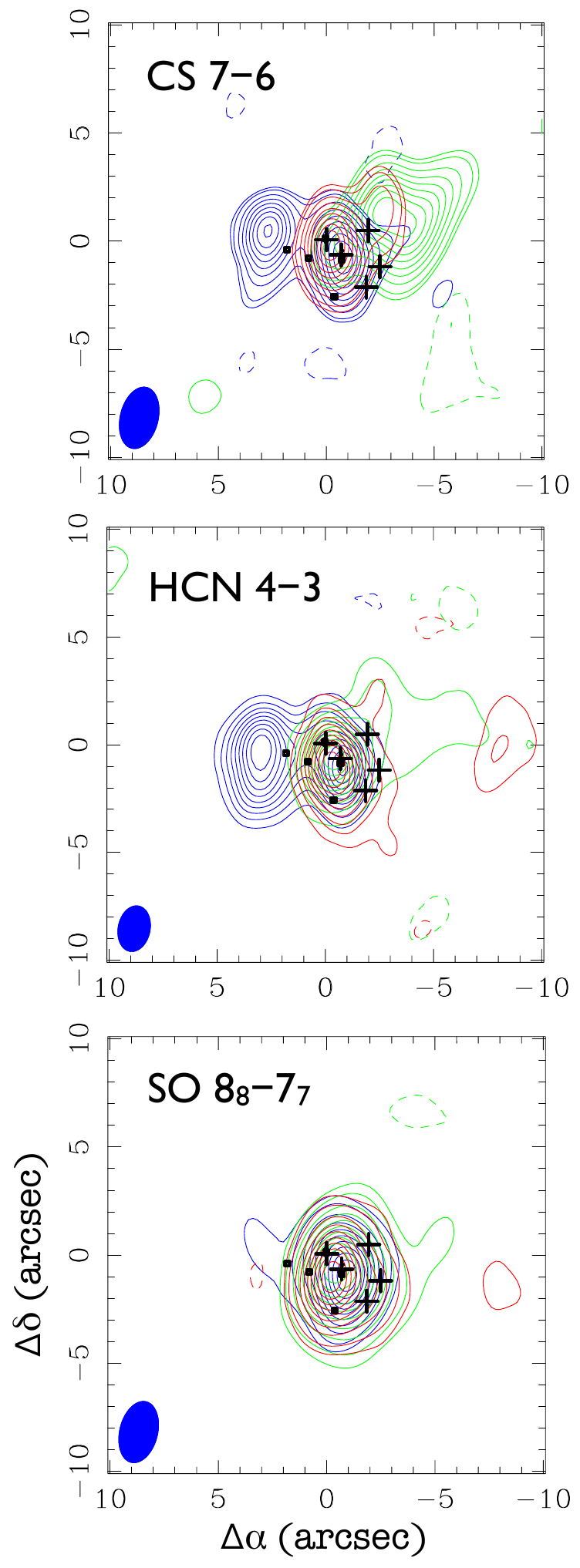}
   \caption{Integrated intensity maps at different velocity ranges
     (blue: $-$50 to $-$43 km s$^{-1}$; green: $-$43 to $-$35 km
     s$^{-1}$; red: $-$35 to $-$28 km s$^{-1}$). We adopted the
     combined-configuration dataset in natural weighting in order to
     show the additional weak emission features. The contour levels of
     each Doppler-shifted component for CS, HCN and SO are 15, 20, 30,
     40,...\%, 10, 20, 30,...\% and 5, 10, 20, 30,...\% of the
     emission peak, respectively. We note that the first contours are
     all $\ge3$ $\sigma$ noise level. The markers are identical to the
     ones in Fig. \ref{fig:cont_map_vis}.}
   \label{fig:CS_HCN_SO_three_color_maps}
\end{figure}

In our data, only the CS $7-6$, HCN $4-3$ and SO $N_J$=$8_8$--$7_7$
emission show a wide velocity range (about 30--40 km s$^{-1}$;
Fig. \ref{fig:mol_hanspec}), presumably due to outflow activity in W3
IRS5.  To study the distribution of the high velocity components
(Fig. \ref{fig:CS_HCN_SO_three_color_maps}), we integrate the emission
over three different velocity ranges (blue: $-$50 to $-$43 km
s$^{-1}$, green: $-$43 to $-$35 km s$^{-1}$ and red: $-$35 to $-$28 km
s$^{-1}$). We took the combined-configuration dataset and used natural
weighting for the analysis. The images are plotted with contours
starting well above $3\sigma$ to minimize the impact of sidelobes,
especially for the green component. As seen in
Fig. \ref{fig:CS_HCN_SO_three_color_maps}, the high-velocity
components of all three molecules peak near zone A and close to SMM2,
suggesting a bipolar outflow with an inclination close to the line of
sight. The existence of a line-of-sight outflow also implies a
rotating structure in the plane of the sky, in which SMM1 and SMM2
form a binary system. Therefore, the small scale velocity gradients
(Fig. \ref{fig:32SO2_HCNv2_vgrad}) seen in HCN 4--3 $v_2=1$ and SO$_2$
may highlight the kinematics of the envelope. Additional blue-shifted
components are seen toward zone D. The green component of CS
highlights the gas in zone C. In the HCN map, the emission in zone D
together with the emission peaking near the offset position
($-8\arcsec$, $0\arcsec$) may form another bipolar outflow in the E--W
direction since the line connecting these peaks passes through zone A
which contains star-forming cores. In this case, the one-arcsecond
scale velocity gradients seen in Fig. \ref{fig:32SO2_HCNv2_vgrad}
imply a complex motion (rotation plus expansion or contraction) in the
common envelope of SMM1 and SMM2. To summarize, the kinematics in W3
IRS5 is very complex and requires additional observations at high
resolution with good image fidelity for further analysis.

\section{Discussion}
\subsection{Star formation in W3 IRS5}
\subsubsection{Different evolutionary stages within W3 IRS5}
Based on the continuum image at 353.6 GHz, we identified five compact
sources (SMM1 to SMM5) toward W3 IRS5 (Fig. \ref{fig:cont_map_vis} and
Table \ref{tab:cont_property}). SMM1 and SMM2 also show compact
cm-wave emission \citep{Wilson03,vanderTak05} indicating that stars
sufficiently massive to ionize their surroundings have formed ($>$8
$M_{\sun}$). The non-detection of compact cm-wave emission toward SMM3,
SMM4 and SMM5 implies that currently those cores have not (yet) formed
massive stars. By comparing the 353.6-GHz continuum with the NICMOS
2.22 $\mu$m emission and cm-wave emission
\citep[Fig. \ref{fig:cont_map_vis}(a),][]{Megeath05,Wilson03,vanderTak05},
we suggest that the eastern part (SMM1 and SMM2) of the source is more
evolved than the western part (SMM3, SMM4 and SMM5). The small core mass ($\sim0.3$ $M_{\sun}$) suggests that SMM3--5 may be forming low-mass stars or are starless cores.

For a virialized star-forming core ($M_{\rm core}+M_{\rm star} \approx M_{\rm vir}$), the 1D virial velocity dispersion ($\sigma_{\rm vir}$; c.f. \citet{McKee07}) is 
\begin{equation} 
\sigma_{\rm vir} = \sqrt{\frac{GM_{\rm vir}}{5r}}, 
\end{equation}	
where $r$ is the radius of the source, $G$ is the gravitational constant, and $M_{\rm vir}$ is the virial mass. Assuming a core radius of $0.5\arcsec$, core gas masses of $\sim$0.3 $M_{\sun}$, and stellar masses of SMM3--5 less than 8 $M_{\sun}$, the 1D virial velocity dispersions should be no more than 1.3 km s$^{-1}$. From our observations of CH$_3$OH line toward SMM3/4 and CS line toward SMM5, the inferred 1D velocity dispersions ($\sigma$ = $\Delta V$/$\sqrt{8\ln{2}}$) are 1.0 km s$^{-1}$ and 1.4 km s$^{-1}$. The large observed line widths suggest that either these cores contain stars just under 8 solar masses, or the clumps may have smaller mass stars, and the large velocities may be the influence of outflows from SMM1/2. In the latter, case, it is not clear if the gas in the cores is gravitationally bound. Subarcsecond observations of dust continuum and outflow tracers should tell if SMM3--5 are low-mass starless leftover cores after the formation of SMM1--2 or harbor low-mass protostars under the influence of outflows from SMM1--2.

\subsubsection{Fragmentation and Jeans analysis}			
From the continuum analysis, we found the core masses of the identified SMM sources are small (0.2--0.6 $M_{\sun}$) and the projected distance between the SMM sources are about $1\arcsec-2\arcsec$ (1800 AU -- 3600 AU). This observed structure, in the meantime, is still embedded in a large massive core with size about $1\arcmin$ and mass of several hundred $M_{\sun}$ \citep[W3-SMS1 in the SCUBA 850 $\mu$m image,][]{DiFrancesco08,YWang12}. At $1\arcmin$ scale, \citet{Megeath96} found that there are $\sim 300$ low-mass stars in the core surrounding W3 IRS5 with a stellar density of few 1000 pc$^{-3}$. Therefore, it is interesting to investigate the cloud structure from $1\arcmin$ scale to $1\arcsec$ to see what determines the overall core/stellar mass distribution and their spatial distribution. To study this question, we perform Jeans analysis in W3 IRS5 based on our SMA continuum image and the JCMT SCUBA 850$\mu$m image \citep{DiFrancesco08,YWang12}. The Jeans length and Jeans mass are expressed as \citep[c.f.][Eq. (9.23) and (9.24)]{Stahler05}:
\begin{equation} 
\lambda_J \approx 0.19\ (\frac{T}{10 {\rm \ K}})^{1/2}\ (\frac{n_{\rm H_2}}{10^4\ {\rm cm^{-3}}})^{-1/2}\ {\rm pc,} 
\end{equation}			
and
\begin{equation} 
M_J \approx 1.0\ (\frac{T}{10 {\rm \ K}})^{3/2}\ (\frac{n_{\rm H_2}}{10^4\ {\rm cm^{-3}}})^{-1/2}\ {\rm M_{\sun},} 
\end{equation}						
where $T$ is the kinetic temperature and $n_{\rm H_2}$ is the H$_2$
volume density. These two quantities are more sensitive to the adopted H$_2$ volume density. To estimate the $1\arcmin$ scale H$_2$ density, we measured the total flux density in a $1\arcmin$ by $1\arcmin$ box centered toward W3 IRS5 using the SCUBA 850 $\mu$m image. Using the same prescription outlined in Sect. 4.1, a total flux density of 160 Jy corresponds to a mass of 950 $M_{\sun}$ assuming a dust temperature of 40 K. The mean H$_2$ volume density is $2.4\times10^{5}$ cm$^{-3}$. Therefore, at $1\arcmin$ we derived the Jeans length and Jeans mass to be 16000 AU ($\sim$9\arcsec) and 2 $M_{\sun}$, respectively. Interestingly, these two characteristic quantities seem to match the observed spatial separations of low-mass stars \citep[18000--12000 AU,][]{Megeath96}, implying that the spacing of the young low mass stars are consistent with gravitational fragmentation. The Jeans masses are greater than the mean stellar mass for a standard initial mass function (0.5 $M_{\sun}$), which is expected given a typical efficiency of low-mass star formation $\sim 30\%$ \citep{Alves07}. Toward the central regions of W3 IRS5, the temperature ($\sim$140 K) and density (few $10^{7}$ cm$^{-3}$) are higher than the values on the scale of the entire core. The corresponding Jeans length and Jeans mass are 2700--3800 AU ($1.5\arcsec$--2.1$\arcsec$) and $\sim$1 $M_{\sun}$, respectively. The Jeans length is consistent with the observed spacing of the SMM cores, implying that gravitational fragmentation may be occurring in the dense, warm center of the cluster.  On the other hands, the implied Jeans masses are much smaller. This immediately leads to the question how the massive stars in SMM1/2 accreted their current masses.

\subsubsection{Accretion rates}	
The largest accretion rate that an object can sustain is essentially given by its mass divided by its free-fall time,
\begin{equation} 
\dot{M}_{\rm star} \approx M_{\rm core}/t_{\rm ff}, 
\end{equation}	
where $t_{\rm ff} = \sqrt{3\pi/32G\bar{\rho}}$ and $\bar{\rho}$ is the mean
density. The masses of the SMM1--5 cores come from Table \ref{tab:cont_property}. We use the observed size for the SMM1--5 cores, which are set to $1\arcsec$ in diameter based on Fig. \ref{fig:cont_map_vis} (a). We adopt stellar masses of 20 $M_{\sun}$ for SMM1 and SMM2 based on the total luminosity of $2\times10^5$ $L_{\sun}$. We use the combined stellar and core masses and the core radii to determine the average baryonic density; the corresponding free-fall times are $\sim$ 1000 yr for SMM1/2, and $>1700$ yr for SMM3--5. These imply mass accretion rates of $5\times10^{-4}$ $M_{\sun}$~yr$^{-1}$ for SMM1/2, and $<2\times10^{-4}$ $M_{\sun}$~yr$^{-1}$ for SMM3--5. The upper limits for the latter follow from the upper limit on the embedded stellar masses of 8 $M_{\sun}$. The accretion rates are consistent with those expected for the formation of massive stars \citep{Keto03}. The combination of small core masses and high accretion rates imply very short time scales of a few thousand years. If the massive stars in SMM1/2 form from accretion of the local cores ($1\arcsec$ scales), the low core masses imply that they are now in a stage that most of core mass has been accrete onto the stellar components. On the other hand, the SMM cores may be accreting material that have been resolved out by the interferometers on scales $>1\arcsec$.  In this case, the reservoir of material from which the cores are accreting is larger than the observed projected spacing of the protostars, implying that the accretion is coming from a global collapse of the molecular core in which the proto-Trapezium is embedded.  

To determine whether it is possible that the massive stars in W3 IRS5 are being fed from the collapse of the surrounding core, we estimate the potential accretion rates from this accretion.  The free-fall time for the central $1\arcmin$ W3--SMS1 condensation, with a total mass of 950 $M_{\sun}$ and an embedded stellar mass of $\lesssim
50$ $M_{\sun}$, is 72000 yr and the free-fall mass accretion rate is $1\times 10^{-2}$ $M_{\sun}$~yr$^{-1}$. We note that the derived accretion rate is the maximum rate of infall assuming free-fall collapse. In the central region, the amount of gas accreted by each star is given by
\begin{equation}
\dot{M} = \rho\pi R^{2}_{\rm acc}v_{\rm inf}
\end{equation}
\citep{Bonnell01}, where $v_{\rm inf}$ is the infall gas velocity, $\rho$ is the density of the gas surrounding the proto-Trapezium, and $R_{\rm acc}$ is the radius within which the infalling gas is accreted onto a specific star. We assume $n_{\rm H_{2}} = 10^6$~cm$^{-3}$ to compute $\rho$, which is intermediate between the density of the surrounding $1\arcmin$ diameter core ($10^5$~cm$^{-3}$) and the core of the inner $1\arcsec$ ($10^7$~cm$^{-3}$). Given FWHM line widths of 2--5 km s$^{-1}$, we adopt a $v_{\rm inf}$ equal to the 3D gas velocity dispersion  implied by the FWHM line width ($\sqrt{3}\Delta V/\sqrt{8\ln{2}}$), which is 1.5--3.7 km s$^{-1}$ with a mean value of 2.6 km s$^{-1}$. The uncertainty of $v_{\rm inf}$ gives a factor of $<3$ in error of accretion rate. The possible values for $R_{\rm acc}$ are the radius set by the the motions of the stars through the gas (Bondi-Hoyle radius) and the radius set by tidal interaction of the stars with the gravitational potential of the cluster core (tidal radius). The actual accretion rate is given by the minimum of these two radii \citep{Bonnell01}. To estimate $R_{\rm acc}$ due to tidal interactions, we use the equation from \citet{Bonnell01} and rewrite in terms of the gas density around the star,
\begin{equation}
R_{\rm acc}^{\rm tidal} = 0.5(\frac{M_{\rm star}}{4/3\pi\rho})^{1/3}.
\end{equation}
In this case, $R_{\rm acc}^{\rm tidal}$ are 4700 AU and 3500 AU for a 20 $M_{\sun}$ and 8 $M_{\sun}$ central star embedded in a core with gas density 10$^{6}$ cm$^{-3}$, respectively. The accretion rates are then $2\times10^{-4}$ $M_{\sun}$~yr$^{-1}$ and $1\times10^{-4}$ $M_{\sun}$~yr$^{-1}$ for the 20 $M_{\sun}$ and 8 $M_{\sun}$ stars, respectively. Alternatively, the motions of the stars through the cloud can limit the gas accretion; only the gas within a radius where the escape velocity is less than the velocity of the star relative to the gas can be accreted. The $R_{\rm acc}$ in this case can be approximated by the Bondi-Hoyle radius:
\begin{equation} 
R_{\rm acc}^{\rm BH} \approx \frac{2GM_{\rm star}}{\sigma^2_{\rm 3D}}. 
\end{equation}
The resulting $R_{\rm acc}^{\rm BH}$ are 5200 AU and 2100 AU for the 20 and 8~$M_{\sun}$ stars, and the mass infall rates are $3\times10^{-4}$ $M_{\sun}$ yr$^{-1}$ for a stellar mass of 20~M$_{\sun}$ and $4\times10^{-5}$ $M_{\sun}$~yr$^{-1}$ for a stellar mass of 8~M$_{\sun}$. 

As a result, from our basic analysis, global collapse of the core and subsequent accrete of the material from the collapse onto the stars can sustain infall rates in excess of $10^{-4}$~$M_{\sun}$~yr$^{-1}$ for SMM1/2 and are sufficient to sustain high mass star formation. For SMM3--5, the rates are 10 times lower. However, the local density around SMM3--5 is closer to $10^{7}$ cm$^{-3}$, the actual densities and the accretion rates may be higher. The fact that the stellar accretion rates is two orders of magnitude less than the free-fall accretion rate for the large cores suggests that there is sufficient infall of material to feed the accretion of the individual stars, even if the actual collapse times is ten free fall times.  

In summary, we find two alternatives.  Either the massive stars accrete locally from their local cores; in this case the small core masses imply W3 IRS5 is at the very end stages (1000 yr) of infall and accretion.  Alternatively, the stars are accreting from the global collapse of a massive, cluster forming core.  This later scenario is similar to the competitive accretion models of massive star formation \citep[e.g.,][]{Bonnell06}. For the two massive objects SMM1/2, the observed densities and velocity widths are consistent with those needed for the global collapse to feed accretion onto the massive stars at rates of $\sim10^{-4}$ $M_{\sun}$ yr$^{-1}$.  For the remaining objects, SMM 3--5, a lower infall rate is estimated; perhaps these are forming intermediate mass stars, although the rate of accretion may rise as they increase in mass as precede by the models of competitive accretion. However, a clear detection of the infall of gas has not yet been observed. This makes the W3 IRS5 cluster a perfect region to further test the competitive accretion model, if the kinematics of the gas motions can be traced on scales from $1\arcmin$ down to $<1\arcsec$. Future single-dish and interferometric observations of this region, e.g., combining SMA or PdBI observations with IRAM 30m measurements, are required to fully map the gas flow across these scales.

\subsection{The molecular environment of W3 IRS5}		
\subsubsection{Distinct molecular zones}	
A JCMT spectral line survey characterized W3 IRS5 as a remarkable
massive star-forming region with rich spectral features from
sulfur-bearing molecules
\citep{Helmich94,Helmich97}. \citet{Helmich94} also reported that the
CH$_3$OH abundance is remarkably low, even compared to typical dark
clouds. From our SMA data, and the results from \citet{YWang12}, we
find that sulfur-bearing molecules such as SO, SO$_2$ and their
isotopologues, strongly peak toward the submillimeter sources SMM1 and
SMM2 (zone A) with extended emission toward SMM3, SMM4 and SMM5
(Fig. \ref{fig:line_mom0}).  This suggests that SMM1 and SMM2 are the
driving center of the sulfur chemistry. Interestingly, typical
hot-core molecules such as CH$_3$OH (Fig. \ref{fig:line_mom0} (k)) and
CH$_3$CN \citep{YWang12} peak toward SMM3/4 (zone B) exclusively,
underlining the chemical differences between zones A and B. Moreover,
toward zone B, there are two distinct regions along the line of sight
with different physical characteristics. The narrow FWHM line widths
seen in CH$_3$OH (2.3 km s$^{-1}$; Table \ref{tab:line_gaussian}) and
CH$_3$CN \citep[1.6 km s$^{-1}$;][]{YWang12} suggest that this `hot
core' source is less turbulent compared to the gas component traced by
SO and SO$_2$, which have larger line widths ($\sim4-7$ km s$^{-1}$;
Table \ref{tab:line_gaussian}). As implied by the excitation analysis
(Sect. 4.2.2), this `hot core' is less dense ($\sim10^{5}$ cm$^{-3}$)
than the gas traced by $^{34}$SO$_2$ ($\ge10^{9}$ cm$^{-3}$). We suggest
that this `hot core' may created by feedback from the star-forming
activity associated with SMM1 and SMM2. It is, however, hard to tell
if this core is heated internally or externally since we do not find
significant temperature differences between $^{34}$SO$_2$ and CH$_3$CN
(Sect. 4.2.2). In addition, there is another interesting chemical
signature in W3 IRS5, where SO and SO$_2$ are peaked toward zone A,
while CS and H$_2$CS are peaked toward zone C
(Fig. \ref{fig:line_mom0}). Although these species are chemically
related \citep[e.g.][]{Charnley97}, this spatial de-correlation
implies that different chemical processes are involved, such as hot
core versus shock chemistry \citep{Hatchell02}.
		
\subsubsection{Abundances of SO, SO$_2$ and CH$_3$OH}	
It has been proposed that the abundance ratios of sulfur-bearing
species can be used to measure the time elapsed since the start of
ice-mantle evaporation in star-forming cores \emph{if} proper physical
conditions are applied
\citep[e.g.][]{Charnley97,vanderTak03,Wakelam04}. Unfortunately, the
rough estimates of the SO and SO$_2$ abundances provided by our data
limit us to a qualitative comparison only. SO and SO$_{2}$ are abundant
toward W3~IRS5. The fractional abundance of SO and SO$_{2}$ (Table
\ref{tab:excitation}) are, respectively, $8.6\times10^{-6}$ and
$9.4\times10^{-7}$ toward zone A, and $5.4\times10^{-6}$ and
$2.8\times10^{-7}$ toward zone B. This corresponds to SO$_2$/SO ratios
of 0.1 and 0.05 for zones A and B, respectively. The uncertainty of
these ratios are factors of few, and SO$_2$ is clearly less abundant
than SO toward W3 IRS5.
			
\citet{Charnley97}, \citet{Hatchell98} and \citet{Wakelam04} present
chemical models that relate the SO$_2$/SO ratio to the time since the
onset of grain-mantle evaporation. Although these models are very
sensitive to the initial mantle composition, cosmic ray ionization
rate, density and temperature, these models show that the evolution of
the fractional abundances of SO and SO$_2$ peak after roughly
$10^{4}$-$10^{5}$ yr after grain-mantle evaporation. Such time scales
are consistent with the free-fall time for the entire cluster
calculated in Sect. 5.1.3 of $(0.8$--$1.4)\times 10^5$ yr.

Toward the emission peak in zone B, we derive a CH$_3$OH abundance of
$5.1\times10^{-8}$, lower by a factor of $10-100$ compared to the
abundance found toward the Orion Compact Ridge \citep{Menten88}. This
low CH$_3$OH abundance implies that the `hot core' toward zone B is
chemically young since large amounts of CH$_3$OH are still
observed in the solid state \citep{Allamandola92}. Since zone B does
show emission from SO and SO$_2$, which is indicative of outflow
activity (Fig. \ref{fig:CS_HCN_SO_three_color_maps}), it is possible
that the `hot core' has just been turned on by external heating from
zone A.

\subsection{Non-detection of CO$^{+}$: an indication of FUV origin}
Just as much as detected emission can provide information, so can the
non-detection of emission tell us something. Molecular ions and
radicals like CN, NO, CO$^{+}$, SO$^{+}$ and SH$^{+}$, are thought to
trace high energy (far-ultraviolet, FUV, and X-ray) radiation toward
star-forming regions \citep{Sternberg95,Maloney96}. Toward W3 IRS5,
\citet{Stauber07} reported detection of two hyperfine transitions of
CO$^{+}$ $N$=3--2, $F$=5/2--3/2 (353.7413 GHz) and $F$=7/2--5/2
(354.0142 GHz) with the James Clerk Maxwell Telescope (JCMT) at
$14\arcsec$ resolution. They concluded that CO$^{+}$ is present in
FUV-irradiated cavity walls which may be part of the outflows along
the line of sight, similar to AFGL 2591
\citep{Stauber07,Benz07,Bruderer09}. However, the detection of X-ray
emission toward W3 IRS5 \citep{Hofner02} with a luminosity of
$\sim5\times10^{30}$ erg s$^{-1}$ implies that CO$^{+}$ may have an
X-ray origin as well. The geometrical dilution of this X-ray flux down
to the cosmic ray ionization level suggests that the source size of
CO$^{+}$ should be about $0.7\arcsec$ if X-rays are the dominant
source of CO$^+$. Therefore interferometric observations would be
helpful to determine if CO$^{+}$ can have an X-ray origin.
						
\begin{figure}[htbp]
   \centering   
   \includegraphics[width=9cm]{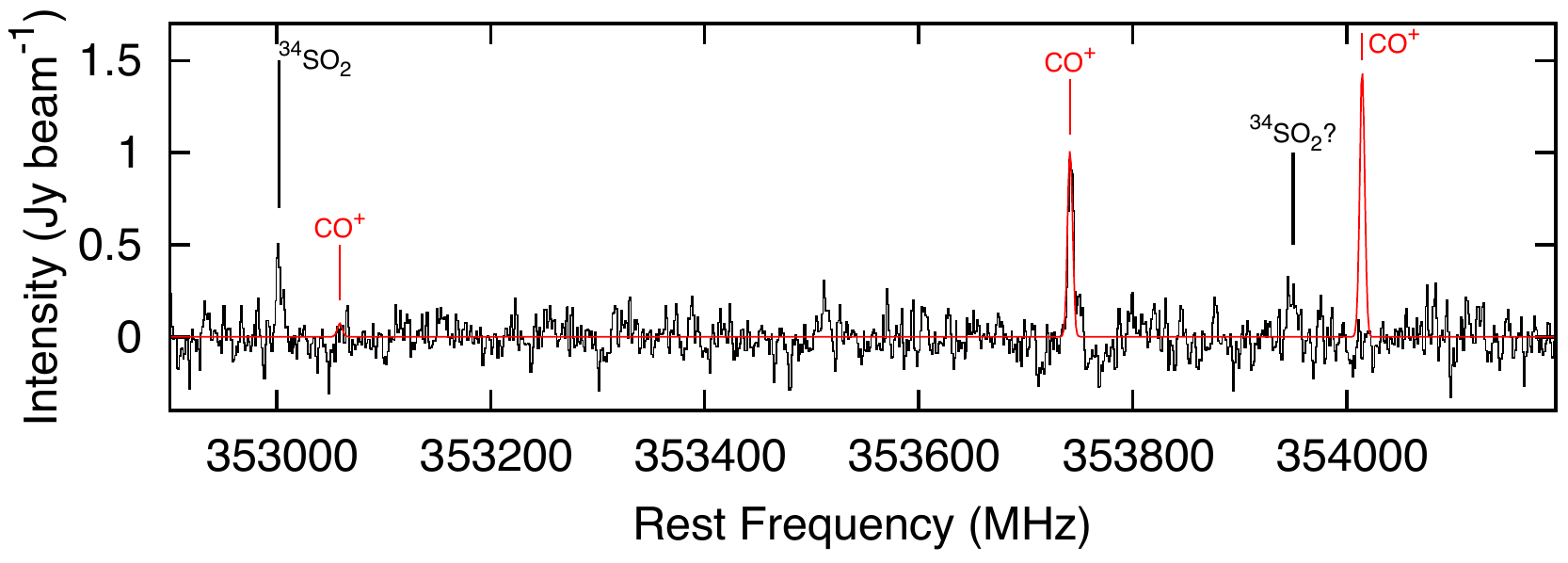}
   \caption{Hanning-smoothed spectrum taken from the
     compact-configuration dataset toward zone A (offset position:
     $-0.42\arcsec$,$-0.43\arcsec$). The black lines are the observed
     data, while the red lines are the model spectrum of CO$^{+}$
     $N$=3--2, $F$=3/2--1/2 (353.0589 GHz), 5/2--3/2 (353.7413 GMHz)
     and 7/2--5/2 (354.0142 GHz) under LTE assumption. The LTE
     intensity ratio of the hyperfine transitions is $\sim$ 1:14:20
     (increasing in frequency). The $F$=7/2--5/2 transition at
     354.0142 GHz is not detected in our SMA data.}
   \label{fig:search_COplus}
\end{figure}

The same set of CO$^{+}$ $N$=3--2 hyperfine transitions are also
covered in our SMA data. Emission at 353.7413 GHz is clearly detected
toward zone A and shows a compact structure
(c.f. Fig. \ref{fig:line_mom0} (d)). However, we do not detect the
transition at 354.0142 GHz (Fig. \ref{fig:search_COplus}). In LTE the
intensity ratio of the CO$^{+}$ $N$=3--2 hyperfine transitions
(including $F$=3/2--1/2 at 353.0589 MHz) is 1:14:20. A similarly
compact source at 354.0142 GHz should therefore have been
detected. Also, the CO$^+$ abundance we derive from the 353.7413 GHz
intensity ($\sim1.3\times10^{-9}$) is much larger than the prediction
from \citet{Stauber07} and \citet{Stauber09} ($\sim 10^{-12}$--$10^{-10}$). We
therefore rule out CO$^+$ as the origin of the emission at 353.7413
GHz.

Instead, we assign the line near 353.741 GHz to $^{33}$SO$_2$
$19_{4,16}-19_{3,17}$ transition, as was suggested by
\citet{Stauber07}. The anomalous CO$^+$ line ratio reported by
\citep{Stauber07} is explained by blending of the CO$^+$ line with
$^{33}$SO$_2$ at 353.741 GHz. If this assignment is correct, our SMA
data should contain three other lines from $^{33}$SO$_2$ (with
hyperfine transitions), near 344.031 GHz ($E_{\rm u}$=397 K), 344.504
GHz ($E_{\rm u}$=232 K) and 354.246 GHz ($E_{\rm u}$=297 K). These
lines are not detected, but they have either smaller line strengths
(344.504 GHz) or decreased intensity due to hyperfine splitting
(344.031, 354.246 GHz). In fact, the detected transitions near 353.741
GHz is expected to be the strongest in the passband. The assignment of
this line to $^{33}$SO$_2$ is further supported by the detection of
its $11_{1,11}-10_{0,10}$ line by \citet{YWang12}. From our data, we
find a column density ratio of $^{32}$SO$_2$/$^{33}$SO$_2$ of
$\sim147$ assuming optically thin lines and LTE conditions (Table
\ref{tab:excitation}). This ratio is consistent with the ISM
$^{32}$S/$^{33}$S ratio of $\sim132$ \citep{Wilson94,Chin96}.

The fact that CO$^+$, detected in the JCMT beam, and undetected in our
SMA data, suggests an extended ($>21\arcsec$) distribution of the
emission consistent with an origin of the emission in the
FUV-irradiated cavity walls, and rules out an X-ray
origin. Alternatively, the non-detection of CO$^{+}$ in our
$\sim2.5\arcsec$ beam may be explained if CO$^{+}$ is destroyed
rapidly with H$_2$ or electrons \citep[][and the references
  therein]{Stauber09} in the high density ($\ge10^{9}$ cm$^{-3}$)
material near SMM1 and SMM2 identified by our data.

\section{Conclusions}
Our conclusions can be summarized as follows.
\begin{enumerate}
  \item The $1\arcsec$ resolution continuum image at 353.6 GHz shows 5
    compact submillimeter sources SMM1--5
    (Fig.\ \ref{fig:cont_map_vis}). SMM1 and SMM2 have embedded
    massive stars ($\sim$20 $M_{\sun}$), judging from the detection of
    compact cm-wave emission. The other sources do not contain massive
    stars currently, and may be forming low-mass stars or simply be
    starless.
  \item The gas densities of SMM1--5 are high as $10^{7}$ cm$^{-3}$, but their \emph{core} masses are surprisingly low, $0.2-0.6$ $M_{\sun}$.
  \item If the massive stars in  W3 IRS5 accrete locally from the surrounding core, the derived small core masses of the SMM sources imply that the main accretion phase is almost concluded. 
  \item From Jeans analysis, the cloud structures from $1\arcmin$ scales down to $1\arcsec$ scales are likely determined by gravitational fragmentation in turbulent environment. 
  \item The free-fall accretion rates toward SMM1 and SMM2 are about
    $5\times10^{-4}$ $M_{\sun}$ yr$^{-1}$, while SMM3--5 have
    accretion rates of less than 1--$2\times10^{-4}$ $M_{\sun}$
    yr$^{-1}$ if their stellar masses are 8 $M_{\sun}$ or
    2--$4\times10^{-5}$ $M_{\sun}$ yr$^{-1}$ if they are starless.
  \item If star formation in W3 IRS5 follows the competitive accretion
    model, global collapse of the core and subsequent accretion of the material from the collapse onto the stars can sustain infall rates in excess of $10^{-4}$~$M_{\sun}$~yr$^{-1}$ for SMM1/2 and are sufficient to sustain high-mass star formation. 
  \item From the molecular line images, we identified four molecular
    zones (Fig. \ref{fig:molecular_zones}). Zone A and B are the major
    places where sulfur chemistry and hot core chemistry are taking
    place. Zone C and D, however, seem kinematically unrelated to zone
    A and B. There is a `hot core' traced by CH$_3$OH and CH$_3$CN
    toward the line of sight of zone B.
  \item The large abundances of SO and SO$_2$ with respect to H$_2$
    (few $10^{-7}$ to $10^{-6}$) derived in zone A and B are
    indicative of active sulfur chemistry.
  \item The low abundance of CH$_3$OH ($5\times10^{-8}$) in the hot
    core toward zone B implies that it may be recently heated either
    internally or externally by feedback from zone A. Zone B
    seems chemically younger than zone A.
  \item The non-detection of CO$^{+}$ $J$=3--2 on small scales
    supports the idea that CO$^+$ is formed in the FUV-irradiated
    outflow cavity walls, and implies that X-rays either do not
    contribute significantly to the CO$^+$ production or that this
    molecule is rapidly destroyed again in the dense material around
    SMM1 and SMM2.
  
\end{enumerate}

\begin{acknowledgements}
We thank the anonymous referee and the editor Malcolm Walmsley for reviewing our paper. We also thank the SMA staff for conducting our observations. The research of K.-S.W. at Leiden Observatory is supported through a
PhD grant from the Nederlandse Onderzoekschool voor Astronomie (NOVA). This research has made use of NASA's Astrophysics Data System Bibliographic Services. 
\end{acknowledgements}

\bibliographystyle{aa} 
\bibliography{reference.bib} 

\begin{thebibliography}{63}
\expandafter\ifx\csname natexlab\endcsname\relax\def\natexlab#1{#1}\fi

\bibitem[{{Abt} \& {Corbally}(2000)}]{Abt00}
{Abt}, H.~A. \& {Corbally}, C.~J. 2000, \apj, 541, 841

\bibitem[{{Allamandola} {et~al.}(1992){Allamandola}, {Sandford}, {Tielens}, \&
  {Herbst}}]{Allamandola92}
{Allamandola}, L.~J., {Sandford}, S.~A., {Tielens}, A.~G.~G.~M., \& {Herbst},
  T.~M. 1992, \apj, 399, 134

\bibitem[{{Alves} {et~al.}(2007){Alves}, {Lombardi}, \& {Lada}}]{Alves07}
{Alves}, J., {Lombardi}, M., \& {Lada}, C.~J. 2007, \aap, 462, L17

\bibitem[{{Benz} {et~al.}(2007){Benz}, {St{\"a}uber}, {Bourke}, {van der Tak},
  {van Dishoeck}, \& {J{\o}rgensen}}]{Benz07}
{Benz}, A.~O., {St{\"a}uber}, P., {Bourke}, T.~L., {et~al.} 2007, \aap, 475,
  549

\bibitem[{{Beuther} {et~al.}(2007){Beuther}, {Churchwell}, {McKee}, \&
  {Tan}}]{Beuther07}
{Beuther}, H., {Churchwell}, E.~B., {McKee}, C.~F., \& {Tan}, J.~C. 2007,
  Protostars and Planets V, 165

\bibitem[{{Bonnell} \& {Bate}(2006)}]{Bonnell06}
{Bonnell}, I.~A. \& {Bate}, M.~R. 2006, \mnras, 370, 488

\bibitem[{{Bonnell} {et~al.}(2001){Bonnell}, {Bate}, {Clarke}, \&
  {Pringle}}]{Bonnell01}
{Bonnell}, I.~A., {Bate}, M.~R., {Clarke}, C.~J., \& {Pringle}, J.~E. 2001,
  \mnras, 323, 785

\bibitem[{{Bruderer} {et~al.}(2009){Bruderer}, {Benz}, {Doty}, {van Dishoeck},
  \& {Bourke}}]{Bruderer09}
{Bruderer}, S., {Benz}, A.~O., {Doty}, S.~D., {van Dishoeck}, E.~F., \&
  {Bourke}, T.~L. 2009, \apj, 700, 872

\bibitem[{{Campbell} {et~al.}(1995){Campbell}, {Butner}, {Harvey}, {Evans},
  {Campbell}, \& {Sabbey}}]{Campbell95}
{Campbell}, M.~F., {Butner}, H.~M., {Harvey}, P.~M., {et~al.} 1995, \apj, 454,
  831

\bibitem[{{Cernicharo} {et~al.}(2011){Cernicharo}, {Spielfiedel}, {Balan{\c
  c}a}, {Dayou}, {Senent}, {Feautrier}, {Faure}, {Cressiot-Vincent},
  {Wiesenfeld}, \& {Pardo}}]{Cernicharo11}
{Cernicharo}, J., {Spielfiedel}, A., {Balan{\c c}a}, C., {et~al.} 2011, \aap,
  531, A103

\bibitem[{{Cesaroni} {et~al.}(1997){Cesaroni}, {Felli}, {Testi}, {Walmsley}, \&
  {Olmi}}]{Cesaroni97}
{Cesaroni}, R., {Felli}, M., {Testi}, L., {Walmsley}, C.~M., \& {Olmi}, L.
  1997, \aap, 325, 725

\bibitem[{{Charnley}(1997)}]{Charnley97}
{Charnley}, S.~B. 1997, \apj, 481, 396

\bibitem[{{Chin} {et~al.}(1996){Chin}, {Henkel}, {Whiteoak}, {Langer}, \&
  {Churchwell}}]{Chin96}
{Chin}, Y.-N., {Henkel}, C., {Whiteoak}, J.~B., {Langer}, N., \& {Churchwell},
  E.~B. 1996, \aap, 305, 960

\bibitem[{{Claussen} {et~al.}(1994){Claussen}, {Gaume}, {Johnston}, \&
  {Wilson}}]{Claussen94}
{Claussen}, M.~J., {Gaume}, R.~A., {Johnston}, K.~J., \& {Wilson}, T.~L. 1994,
  \apjl, 424, L41

\bibitem[{{Di Francesco} {et~al.}(2008){Di Francesco}, {Johnstone}, {Kirk},
  {MacKenzie}, \& {Ledwosinska}}]{DiFrancesco08}
{Di Francesco}, J., {Johnstone}, D., {Kirk}, H., {MacKenzie}, T., \&
  {Ledwosinska}, E. 2008, \apjs, 175, 277

\bibitem[{{Fontani} {et~al.}(2012){Fontani}, {Caselli}, {Zhang}, {Brand},
  {Busquet}, \& {Palau}}]{Fontani12}
{Fontani}, F., {Caselli}, P., {Zhang}, Q., {et~al.} 2012, \aap, 541, A32

\bibitem[{{Gibb} {et~al.}(2007){Gibb}, {Davis}, \& {Moore}}]{Gibb07}
{Gibb}, A.~G., {Davis}, C.~J., \& {Moore}, T.~J.~T. 2007, \mnras, 382, 1213

\bibitem[{{Green}(1986)}]{Green86}
{Green}, S. 1986, \apj, 309, 331

\bibitem[{{Green}(1995)}]{Green95}
{Green}, S. 1995, \apjs, 100, 213

\bibitem[{{Hatchell} {et~al.}(1998){Hatchell}, {Thompson}, {Millar}, \&
  {MacDonald}}]{Hatchell98}
{Hatchell}, J., {Thompson}, M.~A., {Millar}, T.~J., \& {MacDonald}, G.~H. 1998,
  \aap, 338, 713

\bibitem[{{Hatchell} \& {Viti}(2002)}]{Hatchell02}
{Hatchell}, J. \& {Viti}, S. 2002, \aap, 381, L33

\bibitem[{{Helmich} {et~al.}(1994){Helmich}, {Jansen}, {de Graauw},
  {Groesbeck}, \& {van Dishoeck}}]{Helmich94}
{Helmich}, F.~P., {Jansen}, D.~J., {de Graauw}, T., {Groesbeck}, T.~D., \& {van
  Dishoeck}, E.~F. 1994, \aap, 283, 626

\bibitem[{{Helmich} \& {van Dishoeck}(1997)}]{Helmich97}
{Helmich}, F.~P. \& {van Dishoeck}, E.~F. 1997, \aaps, 124, 205

\bibitem[{{Ho} {et~al.}(2004){Ho}, {Moran}, \& {Lo}}]{Ho04}
{Ho}, P.~T.~P., {Moran}, J.~M., \& {Lo}, K.~Y. 2004, \apjl, 616, L1

\bibitem[{{Hofner} {et~al.}(2002){Hofner}, {Delgado}, {Whitney}, {Churchwell},
  \& {Linz}}]{Hofner02}
{Hofner}, P., {Delgado}, H., {Whitney}, B., {Churchwell}, E., \& {Linz}, H.
  2002, \apjl, 579, L95

\bibitem[{{Imai} {et~al.}(2000){Imai}, {Kameya}, {Sasao}, {Miyoshi}, {Deguchi},
  {Horiuchi}, \& {Asaki}}]{Imai00}
{Imai}, H., {Kameya}, O., {Sasao}, T., {et~al.} 2000, \apj, 538, 751

\bibitem[{{Keto}(2003)}]{Keto03}
{Keto}, E. 2003, \apj, 599, 1196

\bibitem[{{Krumholz} \& {Bonnell}(2009)}]{Krumholz09}
{Krumholz}, M.~R. \& {Bonnell}, I.~A. 2009, {Models for the formation of
  massive stars}, ed. G.~{Chabrier} (Cambridge University Press), 288

\bibitem[{{Longmore} {et~al.}(2011){Longmore}, {Pillai}, {Keto}, {Zhang}, \&
  {Qiu}}]{Longmore11}
{Longmore}, S.~N., {Pillai}, T., {Keto}, E., {Zhang}, Q., \& {Qiu}, K. 2011,
  \apj, 726, 97

\bibitem[{{Maloney} {et~al.}(1996){Maloney}, {Hollenbach}, \&
  {Tielens}}]{Maloney96}
{Maloney}, P.~R., {Hollenbach}, D.~J., \& {Tielens}, A.~G.~G.~M. 1996, \apj,
  466, 561

\bibitem[{{McKee} \& {Ostriker}(2007)}]{McKee07}
{McKee}, C.~F. \& {Ostriker}, E.~C. 2007, \araa, 45, 565

\bibitem[{{McKee} \& {Tan}(2003)}]{McKee03}
{McKee}, C.~F. \& {Tan}, J.~C. 2003, \apj, 585, 850

\bibitem[{{Megeath} {et~al.}(1996){Megeath}, {Herter}, {Beichman}, {Gautier},
  {Hester}, {Rayner}, \& {Shupe}}]{Megeath96}
{Megeath}, S.~T., {Herter}, T., {Beichman}, C., {et~al.} 1996, \aap, 307, 775

\bibitem[{{Megeath} {et~al.}(2008){Megeath}, {Townsley}, {Oey}, \&
  {Tieftrunk}}]{Megeath08}
{Megeath}, S.~T., {Townsley}, L.~K., {Oey}, M.~S., \& {Tieftrunk}, A.~R. 2008,
  {Low and High Mass Star Formation in the W3, W4, and W5 Regions}, ed.
  B.~{Reipurth}, 264

\bibitem[{{Megeath} {et~al.}(2005){Megeath}, {Wilson}, \& {Corbin}}]{Megeath05}
{Megeath}, S.~T., {Wilson}, T.~L., \& {Corbin}, M.~R. 2005, \apjl, 622, L141

\bibitem[{{Menten} {et~al.}(1988){Menten}, {Walmsley}, {Henkel}, \&
  {Wilson}}]{Menten88}
{Menten}, K.~M., {Walmsley}, C.~M., {Henkel}, C., \& {Wilson}, T.~L. 1988,
  \aap, 198, 253

\bibitem[{{Mitchell} {et~al.}(1992){Mitchell}, {Hasegawa}, \&
  {Schella}}]{Mitchell92}
{Mitchell}, G.~F., {Hasegawa}, T.~I., \& {Schella}, J. 1992, \apj, 386, 604

\bibitem[{{M{\"u}ller} {et~al.}(2005){M{\"u}ller}, {Schl{\"o}der}, {Stutzki},
  \& {Winnewisser}}]{Muller05}
{M{\"u}ller}, H.~S.~P., {Schl{\"o}der}, F., {Stutzki}, J., \& {Winnewisser}, G.
  2005, Journal of Molecular Structure, 742, 215

\bibitem[{{Ohishi} {et~al.}(1992){Ohishi}, {Irvine}, \& {Kaifu}}]{Ohishi92}
{Ohishi}, M., {Irvine}, W.~M., \& {Kaifu}, N. 1992, in IAU Symposium, Vol. 150,
  Astrochemistry of Cosmic Phenomena, ed. P.~D. {Singh}, 171

\bibitem[{{Ossenkopf} \& {Henning}(1994)}]{Ossenkopf94}
{Ossenkopf}, V. \& {Henning}, T. 1994, \aap, 291, 943

\bibitem[{{Peretto} {et~al.}(2006){Peretto}, {Andr{\'e}}, \&
  {Belloche}}]{Peretto06}
{Peretto}, N., {Andr{\'e}}, P., \& {Belloche}, A. 2006, \aap, 445, 979

\bibitem[{{Pickett} {et~al.}(1998){Pickett}, {Poynter}, {Cohen}, {Delitsky},
  {Pearson}, \& {M{\"u}ller}}]{Pickett98}
{Pickett}, H.~M., {Poynter}, R.~L., {Cohen}, E.~A., {et~al.} 1998, \jqsrt, 60,
  883

\bibitem[{{Pillai} {et~al.}(2011){Pillai}, {Kauffmann}, {Wyrowski}, {Hatchell},
  {Gibb}, \& {Thompson}}]{Pillai11}
{Pillai}, T., {Kauffmann}, J., {Wyrowski}, F., {et~al.} 2011, \aap, 530, A118

\bibitem[{{Pineau des Forets} {et~al.}(1993){Pineau des Forets}, {Roueff},
  {Schilke}, \& {Flower}}]{Pineau-des-Forets93}
{Pineau des Forets}, G., {Roueff}, E., {Schilke}, P., \& {Flower}, D.~R. 1993,
  \mnras, 262, 915

\bibitem[{{Ridge} \& {Moore}(2001)}]{Ridge01}
{Ridge}, N.~A. \& {Moore}, T.~J.~T. 2001, \aap, 378, 495

\bibitem[{{Rod{\'o}n} {et~al.}(2008){Rod{\'o}n}, {Beuther}, {Megeath}, \& {van
  der Tak}}]{Rodon08}
{Rod{\'o}n}, J.~A., {Beuther}, H., {Megeath}, S.~T., \& {van der Tak}, F.~F.~S.
  2008, \aap, 490, 213

\bibitem[{{Sault} {et~al.}(1995){Sault}, {Teuben}, \& {Wright}}]{Sault95}
{Sault}, R.~J., {Teuben}, P.~J., \& {Wright}, M.~C.~H. 1995, in Astronomical
  Society of the Pacific Conference Series, Vol.~77, Astronomical Data Analysis
  Software and Systems IV, ed. R.~A. {Shaw}, H.~E. {Payne}, \& J.~J.~E.
  {Hayes}, 433

\bibitem[{{Sch{\"o}ier} {et~al.}(2005){Sch{\"o}ier}, {van der Tak}, {van
  Dishoeck}, \& {Black}}]{Schoier05}
{Sch{\"o}ier}, F.~L., {van der Tak}, F.~F.~S., {van Dishoeck}, E.~F., \&
  {Black}, J.~H. 2005, \aap, 432, 369

\bibitem[{{Scoville} {et~al.}(1993){Scoville}, {Carlstrom}, {Chandler},
  {Phillips}, {Scott}, {Tilanus}, \& {Wang}}]{Scoville93}
{Scoville}, N.~Z., {Carlstrom}, J.~E., {Chandler}, C.~J., {et~al.} 1993, \pasp,
  105, 1482

\bibitem[{{Spielfiedel} {et~al.}(2009){Spielfiedel}, {Senent}, {Dayou},
  {Balan{\c c}a}, {Cressiot-Vincent}, {Faure}, {Wiesenfeld}, \&
  {Feautrier}}]{Spielfiedel09}
{Spielfiedel}, A., {Senent}, M.-L., {Dayou}, F., {et~al.} 2009, \jcp, 131,
  014305

\bibitem[{{Stahler} \& {Palla}(2005)}]{Stahler05}
{Stahler}, S.~W. \& {Palla}, F. 2005, {The Formation of Stars (Weinheim:
  Wiley-VCH)}

\bibitem[{{St{\"a}uber} {et~al.}(2007){St{\"a}uber}, {Benz}, {J{\o}rgensen},
  {van Dishoeck}, {Doty}, \& {van der Tak}}]{Stauber07}
{St{\"a}uber}, P., {Benz}, A.~O., {J{\o}rgensen}, J.~K., {et~al.} 2007, \aap,
  466, 977

\bibitem[{{St{\"a}uber} \& {Bruderer}(2009)}]{Stauber09}
{St{\"a}uber}, P. \& {Bruderer}, S. 2009, \aap, 505, 195

\bibitem[{{Sternberg} \& {Dalgarno}(1995)}]{Sternberg95}
{Sternberg}, A. \& {Dalgarno}, A. 1995, \apjs, 99, 565

\bibitem[{{Tieftrunk} {et~al.}(1997){Tieftrunk}, {Gaume}, {Claussen}, {Wilson},
  \& {Johnston}}]{Tieftrunk97}
{Tieftrunk}, A.~R., {Gaume}, R.~A., {Claussen}, M.~J., {Wilson}, T.~L., \&
  {Johnston}, K.~J. 1997, \aap, 318, 931

\bibitem[{{van der Tak} {et~al.}(2007){van der Tak}, {Black}, {Sch{\"o}ier},
  {Jansen}, \& {van Dishoeck}}]{vanderTak07}
{van der Tak}, F.~F.~S., {Black}, J.~H., {Sch{\"o}ier}, F.~L., {Jansen}, D.~J.,
  \& {van Dishoeck}, E.~F. 2007, \aap, 468, 627

\bibitem[{{van der Tak} {et~al.}(2003){van der Tak}, {Boonman}, {Braakman}, \&
  {van Dishoeck}}]{vanderTak03}
{van der Tak}, F.~F.~S., {Boonman}, A.~M.~S., {Braakman}, R., \& {van
  Dishoeck}, E.~F. 2003, \aap, 412, 133

\bibitem[{{van der Tak} {et~al.}(2005){van der Tak}, {Tuthill}, \&
  {Danchi}}]{vanderTak05}
{van der Tak}, F.~F.~S., {Tuthill}, P.~G., \& {Danchi}, W.~C. 2005, \aap, 431,
  993

\bibitem[{{Wakelam} {et~al.}(2004){Wakelam}, {Caselli}, {Ceccarelli}, {Herbst},
  \& {Castets}}]{Wakelam04}
{Wakelam}, V., {Caselli}, P., {Ceccarelli}, C., {Herbst}, E., \& {Castets}, A.
  2004, \aap, 422, 159

\bibitem[{{Wang} {et~al.}(2012){Wang}, {Beuther}, {Zhang}, {Bik}, {Rod{\'o}n},
  {Jiang}, \& {Fallscheer}}]{YWang12}
{Wang}, Y., {Beuther}, H., {Zhang}, Q., {et~al.} 2012, \apj, 754, 87

\bibitem[{{Wilson} {et~al.}(2003){Wilson}, {Boboltz}, {Gaume}, \&
  {Megeath}}]{Wilson03}
{Wilson}, T.~L., {Boboltz}, D.~A., {Gaume}, R.~A., \& {Megeath}, S.~T. 2003,
  \apj, 597, 434

\bibitem[{{Wilson} \& {Rood}(1994)}]{Wilson94}
{Wilson}, T.~L. \& {Rood}, R. 1994, \araa, 32, 191

\bibitem[{{Zinnecker} \& {Yorke}(2007)}]{Zinnecker07}
{Zinnecker}, H. \& {Yorke}, H.~W. 2007, \araa, 45, 481

\end{thebibliography}

\end{document}